# The landscape of quantum technology companies in 2026

**Zeki Can Seskir[1]*, Orçun Koçak[2] and Arsev Umur Aydinoglu[3]**

[1] Institute for Technology Assessment and Systems Analysis (ITAS), Karlsruhe Institute of Technology (KIT), Karlsruhe, Germany

[2] Qubitrium A.Ş., İstanbul, Türkiye

[3] Science and Technology Policy Studies, Graduate School of Social Sciences, Middle East Technical University, Ankara, Türkiye

* Correspondence: zeki.seskir@kit.edu

ORCID: Zeki Can Seskir 0000-0001-6779-8676; Arsev Umur Aydinoglu 0000-0001-8857-6001

## Abstract

We update our 2022 mapping of 441 quantum technology (QT) companies and examine how the industry has developed. The principal dataset now includes 1,092 companies founded with a QT focus, of which 935 remain active. Public sources and archived company websites provide evidence on company formation, technology fields and geography, alongside acquisitions, closures, valuations and involvement in artificial intelligence, defence and space. Company formation continued beyond the apparent 2018 peak in the earlier study. Computing software and applications remain the largest field, with suppliers and supporting businesses accounting for almost a third of active companies. Countries with similar company totals can have different technological profiles, and regional comparisons reveal concentrations obscured by country rankings. The first companies generally preceded the selected national policy milestones. Supporting businesses usually followed technical companies in the same jurisdiction. AI is more prevalent than defence in current company activities, while defence shows the larger increase from the earliest assessed evidence. Space relevance is documented for more than a quarter of active companies, often alongside defence involvement.

## 1 Introduction

Quantum technology (QT) companies build quantum computers and sensors, develop communication systems, supply specialised components and help other organisations use these technologies. Lists organised around a particular technology, country or investor portfolio can give very different impressions of the industry. Mapping companies across these boundaries helps us understand who participates in commercial QT and how the industry is developing.

Our 2022 study identified 441 companies in 42 countries and examined their founding years, geographical distribution and technology fields [1]. It found a growing ecosystem with distinct regional profiles and an emerging population of suppliers and supporting businesses. An updated mapping allows us to examine how that population has changed and how it relates to established firms and national initiatives. The apparent 2018 peak in company formation also left open whether formation was slowing or newer companies were still missing from the data. In that study, "start-ups" meant companies founded with QT as their primary business, irrespective of age [1]. The Quantum Economic Development Consortium (QED-C) describes "pure-play" QT companies as those dedicating all their resources to quantum-relevant technologies [2]. Here we simply use "QT companies", retaining our eligibility criteria and distinguishing companies founded with a QT focus from those that entered QT later.

Demonstrations of logical quantum processing [3], satellite quantum key distribution [4], magnetic navigation [5] and clinical measurements using quantum sensors [6] illustrate the different development paths pursued across QT. Commercial participation extends well beyond companies building complete quantum systems. Component suppliers, software developers, consultants and training businesses connect these developments to other industries and potential users. We also examine companies' involvement in artificial intelligence (AI), defence and space.

This paper returns to the three research questions of the earlier study:

RQ1 What are the temporal and geographical distributions of QT companies?

RQ2 How can QT companies be categorised, and how are these categories populated?

RQ3 What patterns can be derived from empirical data on trends?

We retain the six broad technology fields used in 2022 and distinguish companies founded with a QT focus from those that entered QT later. We also separate post-quantum cryptography (PQC), record acquisitions and closures, and compare historical and current involvement in AI, defence and space.

We also revisit seven arguments drawn from the discussion of our earlier study [1, pp. 10–12], labelled here A1 to A7.

A1 Early company activity may precede or catalyse national initiatives, followed by more companies and greater diversity of technology fields.

A2 Established firms may limit opportunities for new quantum-sensing companies.

A3 Growing and diversifying industry needs support the emergence of service companies.

A4 Regional comparisons reveal patterns that country rankings obscure.

A5 Classification may require greater granularity as the ecosystem develops.

A6 Company counts alone cannot represent economic significance.

A7 Nonprofits, consortia and other supporting organisations warrant their own ecosystem study.

We also return to two prospective questions. PQ1 asks whether diversification would continue or consolidation become dominant, and PQ2 whether the apparent 2018 peak in company formation would persist. Section 2 describes the data collection, company classifications and analytical approach. Section 3 presents the results, and Section 4 discusses them in relation to the earlier research questions and arguments. Section 5 concludes.

# 2 Methods

## 2.1 Identifying and selecting companies

We began with the QT Stakeholder Preliminary List [7], which extended our earlier company list and informed a study of QT communities [8], and Qubitrium's internal company repository. We added companies from directories, association memberships, investor and accelerator portfolios, university spin-out lists, publications and policy documents. Collection took place mainly in August and September 2026, with further checks through 2 November. The ninth edition of Understanding Quantum Technologies [9], published in late September, helped us check and extend the collection. In total, 135 sources contributed companies. Mapping their geographical coverage guided further searches in less well-covered countries (Figure 1).

We included identifiable commercial ventures with substantive QT activity: developing computing, communication or sensing technologies, supplying dedicated enabling technologies, or providing services specifically for QT. Diversified incumbents, internal corporate divisions, universities, research groups, public programmes and nonprofit organisations were excluded from the company analysis. Separately operating corporate ventures and joint ventures were assessed individually. Generic photonics, cybersecurity, quantum-related branding or hypothetical applications did not qualify. We applied no company-age limit and retained eligible companies after acquisition, public listing or closure.

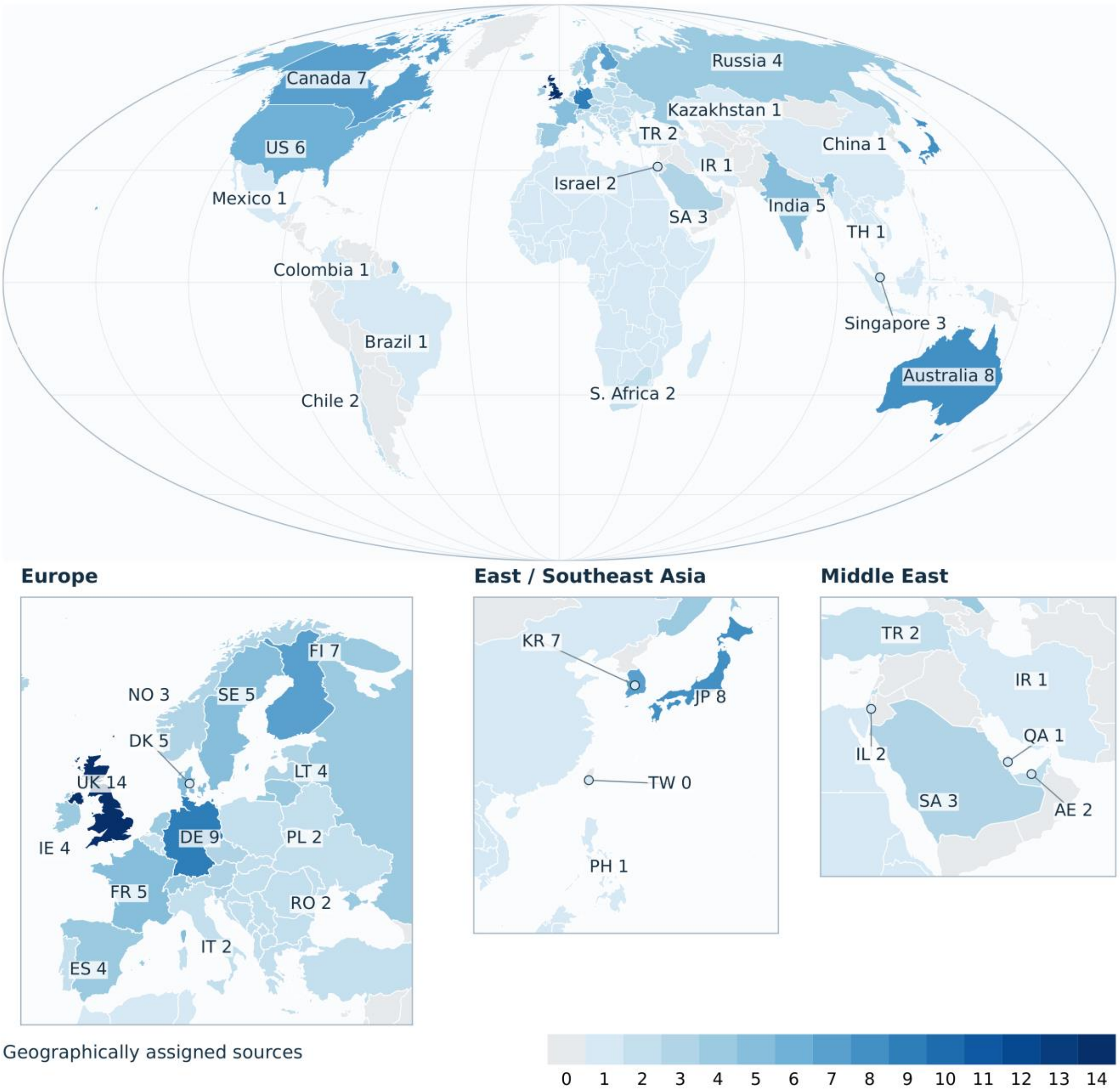

**Figure 1.** Geographical coverage of 135 company-discovery sources. The 108 sources with identifiable country coverage are counted in each country they cover; 27 global or regional sources are not geographically allocated. Shading represents the number of discovery sources covering each country.

After reconciling company names, websites and corporate histories, we retained 1,792 eligible companies (Figure 2). Public listing and corporate origin are recorded as characteristics of these companies, so they do not add to the total.

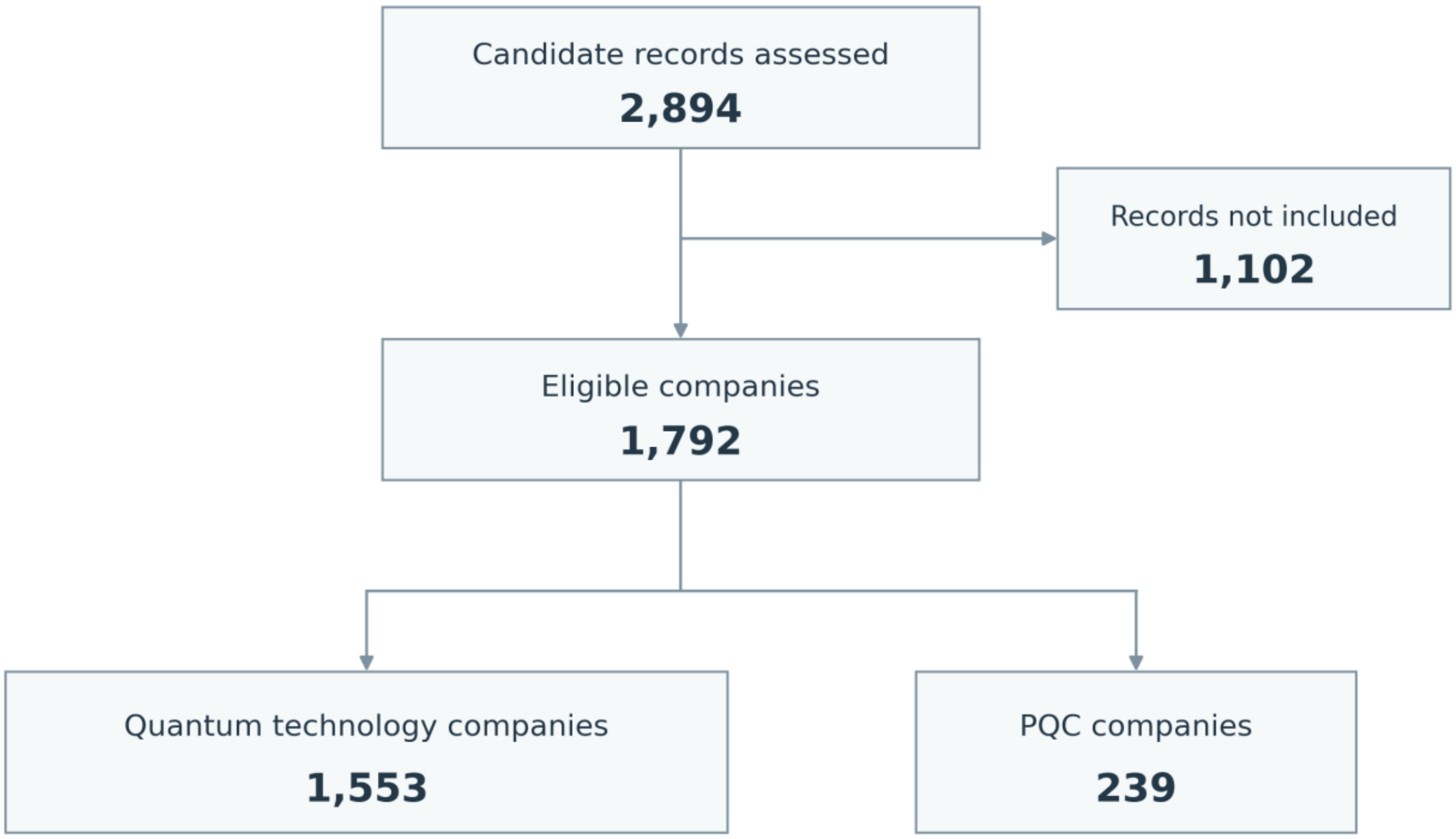

**Figure 2.** Company selection. Screening 2,894 candidate and historical records yielded 1,792 distinct eligible companies: 1,553 with current or historical QT activity and 239 in the separate PQC cohort. The 1,102 records not retained include exclusions, unresolved candidates and consolidated identities. Public listings and corporate origins overlap the primary groups.

OpenAI's GPT-6 Astra model assisted with searches, evidence comparison and proposed classifications. The lead author reviewed the evidence and resolved classification decisions. ChatGPT also supported analysis code, figures and manuscript preparation.

Colleagues working in QT reviewed the dataset and an earlier manuscript. Their feedback prompted further checks of acquisitions and mergers, additions and corrections to the Chinese company population, and revisions to the paper.

We checked acquisitions and mergers against news reports from The Quantum Insider and Quantum Zeitgeist. We count each acquired or merged company once, keeping QT-origin companies, later entrants and PQC companies in their respective groups.

## 2.2 Company histories and classifications

We established founding years and countries of origin from company histories, registers and dated profiles, distinguishing founding from later incorporation or restructuring. Historical activity was assessed using the earliest usable material from the Internet Archive's Wayback Machine, supplemented by contemporary announcements and other dated sources. We checked that these sources referred to the same company. Historical classifications therefore describe the earliest activity we could assess, which may postdate founding. For some recent companies, current material served as a proxy for historical activity.

We distinguish QT-origin companies, founded with a QT focus, from later entrants that developed QT activities after founding. Companies that subsequently left QT are kept separately. The separate PQC group contains PQC-only businesses and those combining PQC with complementary technologies or supporting services. Companies combining PQC with substantive quantum computing, communication or sensing remain in the relevant QT group.

We retain six main fields: computing hardware; computing software and applications; cryptography and communication; sensing and metrology; complementary technologies; and supporting services. Complementary technologies include inputs such as cryogenics and control electronics. Supporting services include consultancy, training and commercial ecosystem services. Each company is assigned one main field. Hardware takes priority over software for full-stack developers, and substantive technical activity takes priority over ancillary consultancy. QT hybrids classified primarily as PQC are grouped with communication; the separate PQC group is excluded.

AI relevance required substantive AI or machine-learning development, functionality or services. Defence relevance required documented defence work, participation in defence programmes or an explicit military market or application. Space relevance required documented involvement in space applications, hardware, services, projects or research. Generic optimisation or supplying AI customers did not establish AI relevance; aerospace positioning, terrestrial free-space optics or agency affiliation alone did not establish space relevance.

We assessed each topic and period separately. Yes denotes qualifying activity; No means none was established from informative material. Possibly indicates suggestive but inconclusive evidence, while Uncertain indicates insufficient evidence. We count a pivot only when a historical No becomes a current Yes. For acquired or inactive companies, current refers to their last recorded activity. Further company checks and evidence reviews informed the final AI, defence and space labels. We manually reviewed random samples of the initial AI, defence, and space classifications for validation checks. Appendix A reports the validation results. The comparative figures use a common Undecided category, combining Possibly and Uncertain for AI, Defence, and Space classifiers.

## 2.3 Populations and comparisons

Current country and field comparisons cover all active QT-origin companies, including those founded in 2026. To examine company formation, we also retain acquired and inactive companies. Temporal comparisons generally end in 2025; where included, the incomplete 2026 cohort is marked. Table 1 summarises the populations used in each analysis.

**Table 1.** Company populations used in the analyses. QT-origin populations include cases assessed as Possibly for historical QT relevance. Public listing and corporate origin are overlapping characteristics.

| Population | Number | Use |
|---|---|---|
| All eligible companies | 1,792 | Full eligible company register |
| QT-origin companies | 1,092 | 935 active, 40 acquired or merged, 117 closed, dissolved or possibly inactive |
| Active QT-origin companies | 935 | Current country, field and regional distributions; all founding years |
| Active QT-origin companies founded through 2025 | 866 | Temporal comparisons; known founding years |
| All QT-origin companies founded through 2025 | 1,022 | Formation, policy and supporting-company analyses; known founding years |
| Space relevance comparison | 975 | Active and acquired or merged QT-origin companies; inactive companies not reviewed |
| Later entrants | 446 | Accepted companies entering QT after founding |
| Separate PQC cohort | 239 | PQC only, or PQC with complementary/supporting activities |
| Companies that left QT | 15 | Added to the historical status comparison |

For comparison with 2022, we retain the earlier regional groups: the United States, EU-27, Australia, Canada and the United Kingdom combined, and the rest of the world. We also compare the United States, EU-27 and China directly. The earlier study included PQC and acquired or inactive companies, while the updated geographical comparisons cover active QT-origin companies and separate PQC.

We compare company formation around selected national policy milestones in 28 countries. These mark the publication, launch or adoption of a strategy or related document, which may follow earlier national support for QT. Twelve countries allow a comparison of the three years before and after the milestone through 2025. Excluding the milestone year, we compare company foundations, represented fields and national shares of worldwide foundations. Appendices A and B provide the documents and supplementary comparisons.

To examine the emergence of supporting businesses, we compare their founding years with those of technical companies in the same jurisdiction. Technical companies comprise the other five fields, including complementary technologies. Appendix A3 provides an additional statistical analysis. Our comparison with established firms draws on Understanding Quantum Technologies [9]. Each company, brand or business-unit entry is counted once in every technology field in which it participates.

The public-company comparison covers companies with confirmed historical and current QT activity that exceeded US$500 million in market capitalisation on 16 October 2026. Market capitalisation is the equity value implied by the share price and outstanding shares. We compare whole-company market capitalisations at the first trading-day close and on 16 October, converting currencies at the exchange rate for each valuation date. Private financing valuations are reported separately, with their dates and valuation bases.

## 2.4 Limitations

Public evidence is uneven across countries, languages and company types. Some recently founded companies may remain unidentified, and changes in activity or closure may go unreported. Finding older companies since the 2022 study shows that discovery delays also affect earlier founding cohorts; we cannot assume a fixed reporting lag. The PQC collection is not a comprehensive survey of cybersecurity companies, and assigning one country and one main field simplifies companies with international operations or several activities, while it omits some operations and the corresponding visibility in our dataset. Historical classifications reflect the public activity visible in the available archive captures. The interval between those observations limits how precisely a pivot can be dated. Companies founded more recently have also had less time to close, be acquired or change focus, so differences between founding cohorts partly reflect their different observation periods.

# 3 Results

## 3.1 Company formation continued beyond the earlier peak

Company formation continued beyond the apparent 2018 peak (Figure 3). Of the 1,022 QT-origin companies with known founding years through 2025, 593 were established during 2021–2025. The largest complete-year cohort is 2025, with 146 companies. A further 66 were founded in 2026, with data collection and checks completed on 2 November 2026.

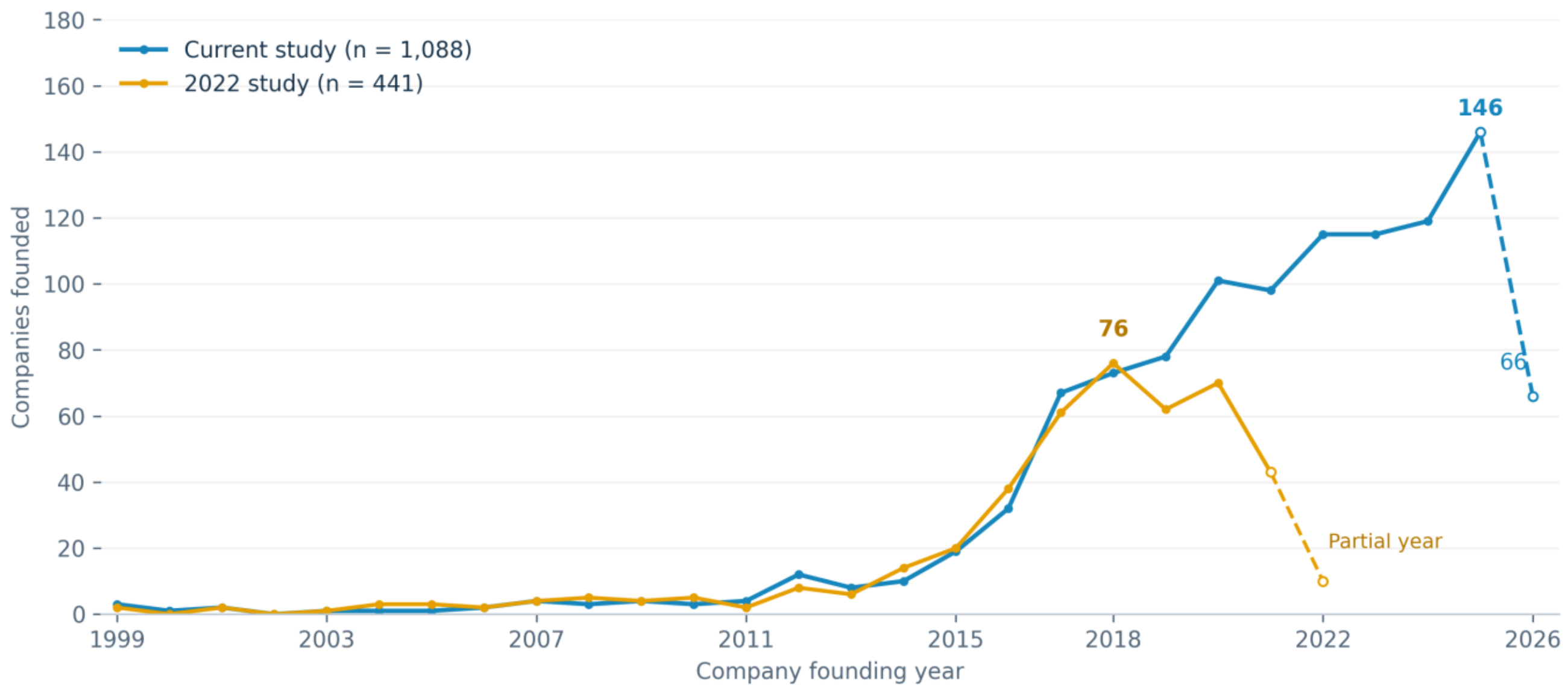

**Figure 3.** Annual company foundations in the 2022 study [1] and the updated QT-origin cohort across active, acquired or merged, and inactive statuses. The updated series contains 1,088 companies with known founding years. Companies that left QT are treated separately. Dashed segments mark incomplete observation years. Coverage and inclusion criteria differ between studies.

Recent company formation spans all six fields (Figure 4). Among companies that remain active, software and applications lead the 2025 cohort with 40 companies, followed by complementary technologies with 27 and supporting businesses with 24. Hardware's largest cohort dates to 2022 and sensing's to 2023. Across fields, the median founding year falls between 2021 and 2022 (Figure S1).

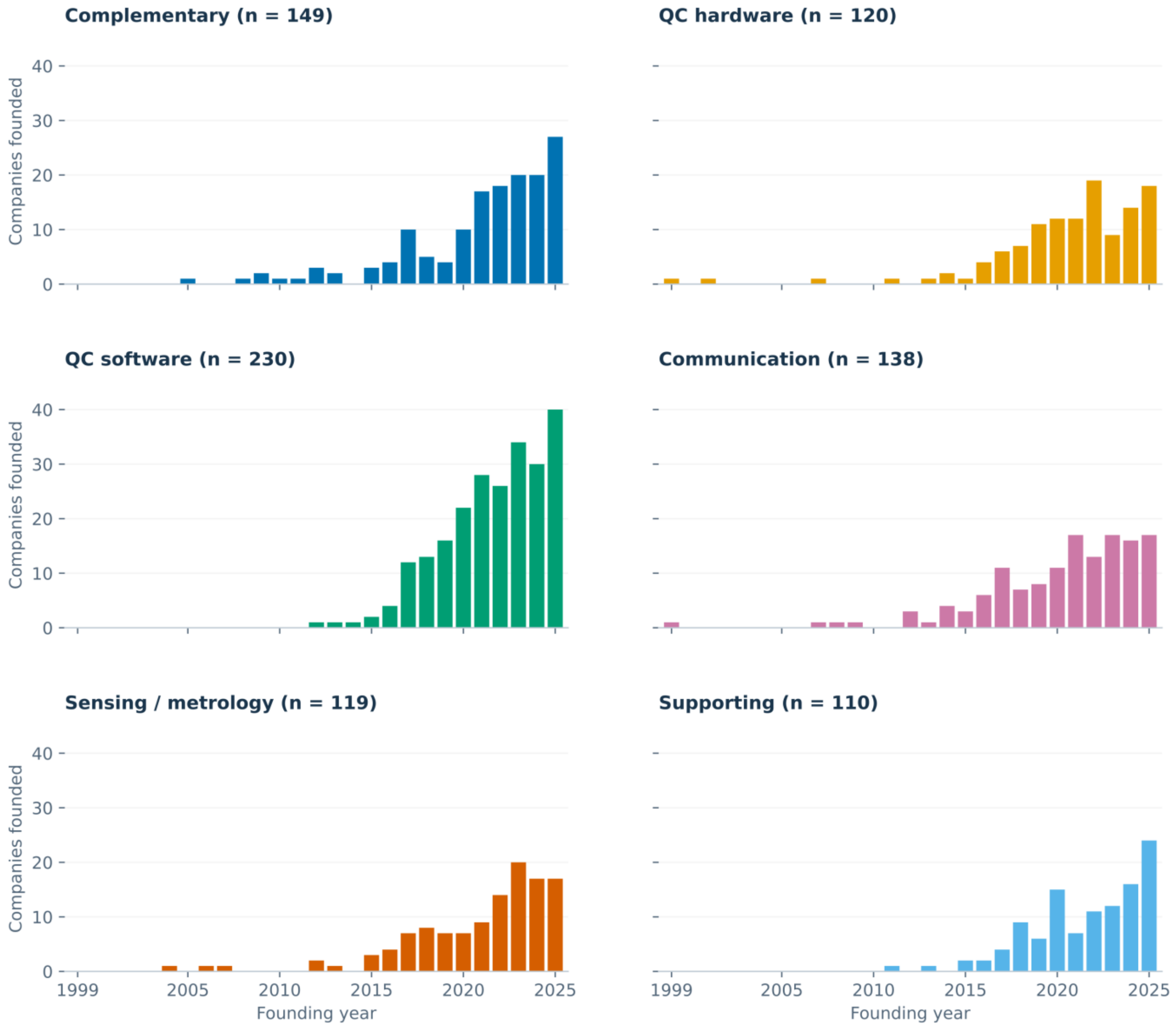

**Figure 4.** Founding years by main technology field for 866 active QT-origin companies founded through 2025. Panels use common axes. QT hybrids carrying a PQC main label are grouped under communication; the separate PQC cohort is excluded.

Among QT-origin companies with known founding years through 2025, 866 remain active, 40 were acquired or merged, and 116 are classified as closed, dissolved or possibly inactive (Figure 5).

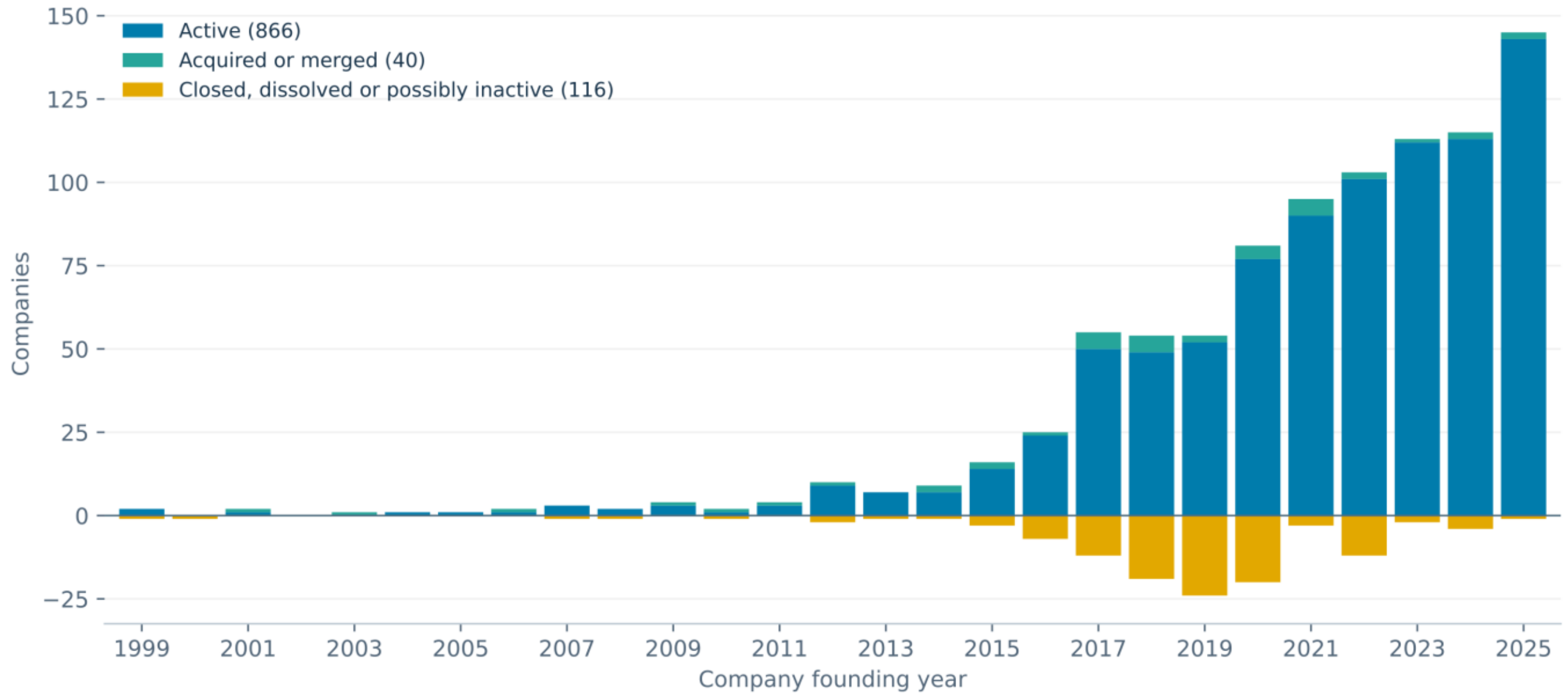

**Figure 5.** Company status by founding year for 1,022 QT-origin companies founded through 2025. Negative bars show closed, dissolved or possibly inactive companies, grouped by founding year. Acquired or merged counts refer to distinct companies.

Closure or inactivity is more common in the 2015–2019 founding cohort than in later cohorts (Figure 6). Of its 275 companies, 65 (23.6%) are closed, dissolved or possibly inactive, compared with 8.5% of companies founded in 2020–2023 and 1.9% in 2024–2025.

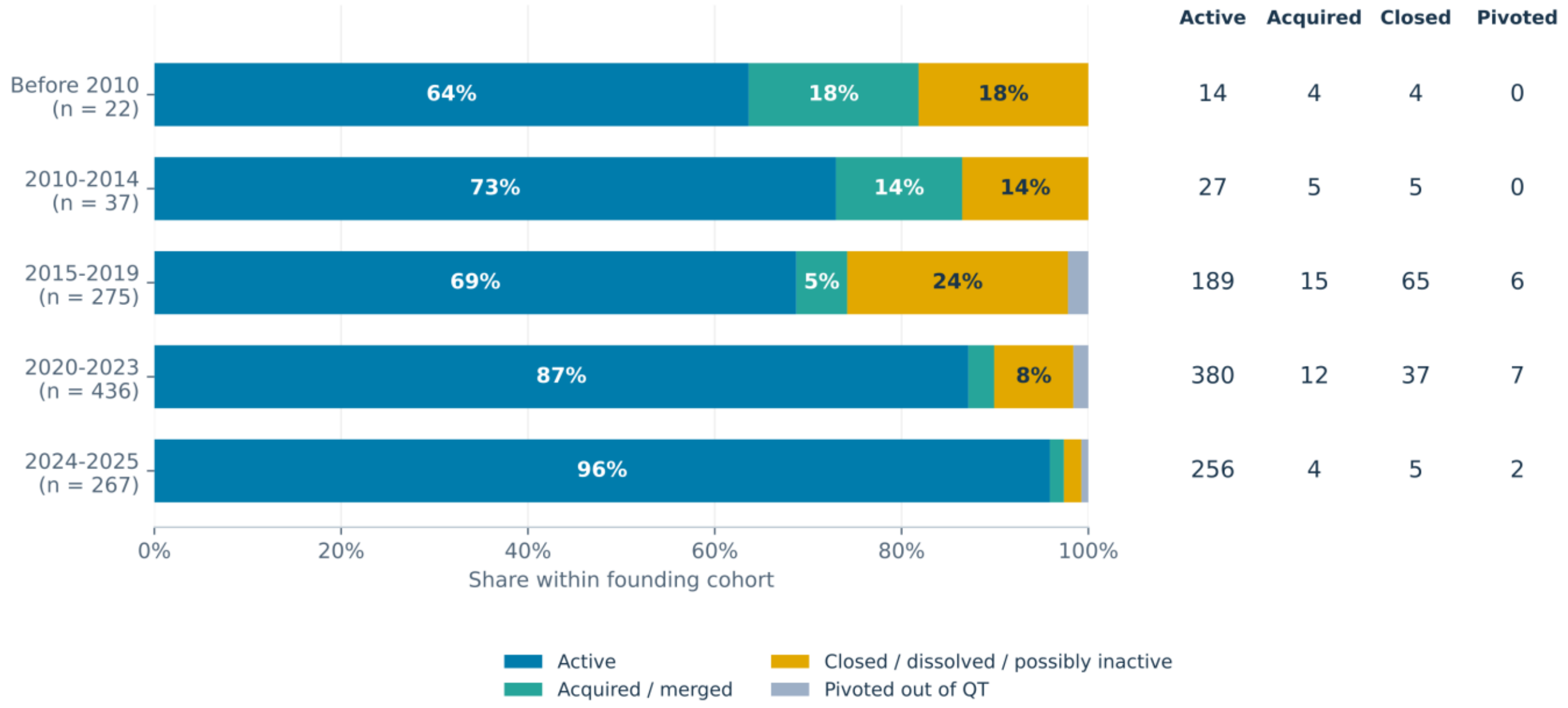

**Figure 6.** Status within founding cohorts through 2025, including 15 companies that left QT (n = 1,037). Percentages use all companies in each cohort. Closed includes dissolved and possibly inactive companies; observation periods differ between cohorts.

### 3.2 A wider geographical landscape with distinct country and regional profiles

Active QT-origin companies are recorded in 72 countries and jurisdictions (Figure 7). The United States leads the country ranking with 189 companies, followed by China (100), Germany (81), the United Kingdom (70) and Canada (58) (Figure S2).

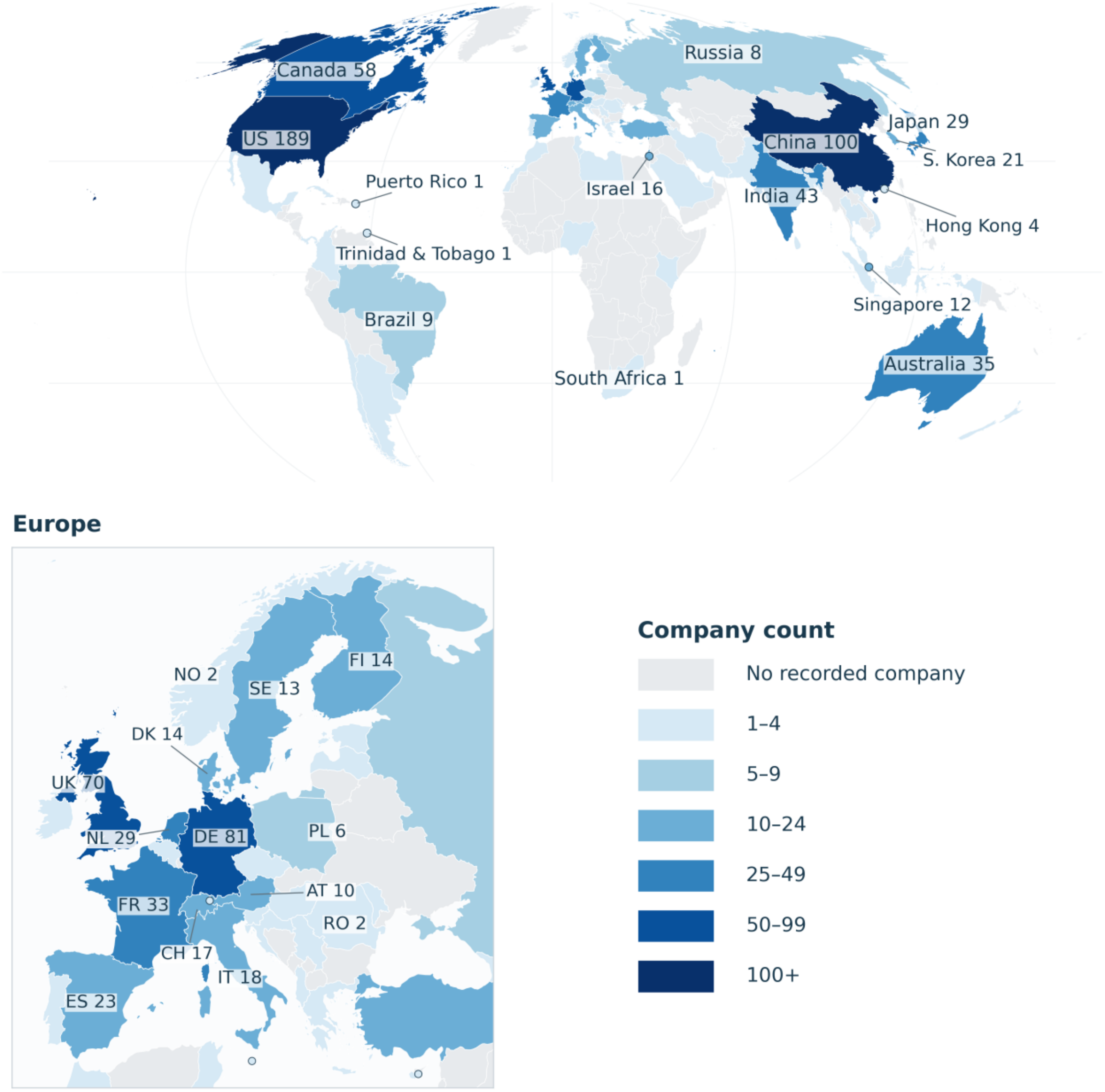

**Figure 7.** Active QT-origin companies by country or jurisdiction (n = 935; 72 jurisdictions), including companies founded in 2026. Each company is assigned once. Small-jurisdiction markers are enlarged for visibility. Boundaries: Natural Earth.

Countries with similar company totals can have different technological profiles (Figure 8). Germany's 81 companies include 24 complementary-technology suppliers, its largest field, while software and applications lead among the United Kingdom's 70 companies, with 22. Japan and the Netherlands each have 29 companies, yet software accounts for 16 in Japan, while complementary technologies lead in the Netherlands with eleven. France’s 33 companies are more evenly spread across fields. Figure S6 presents the within-country percentages.

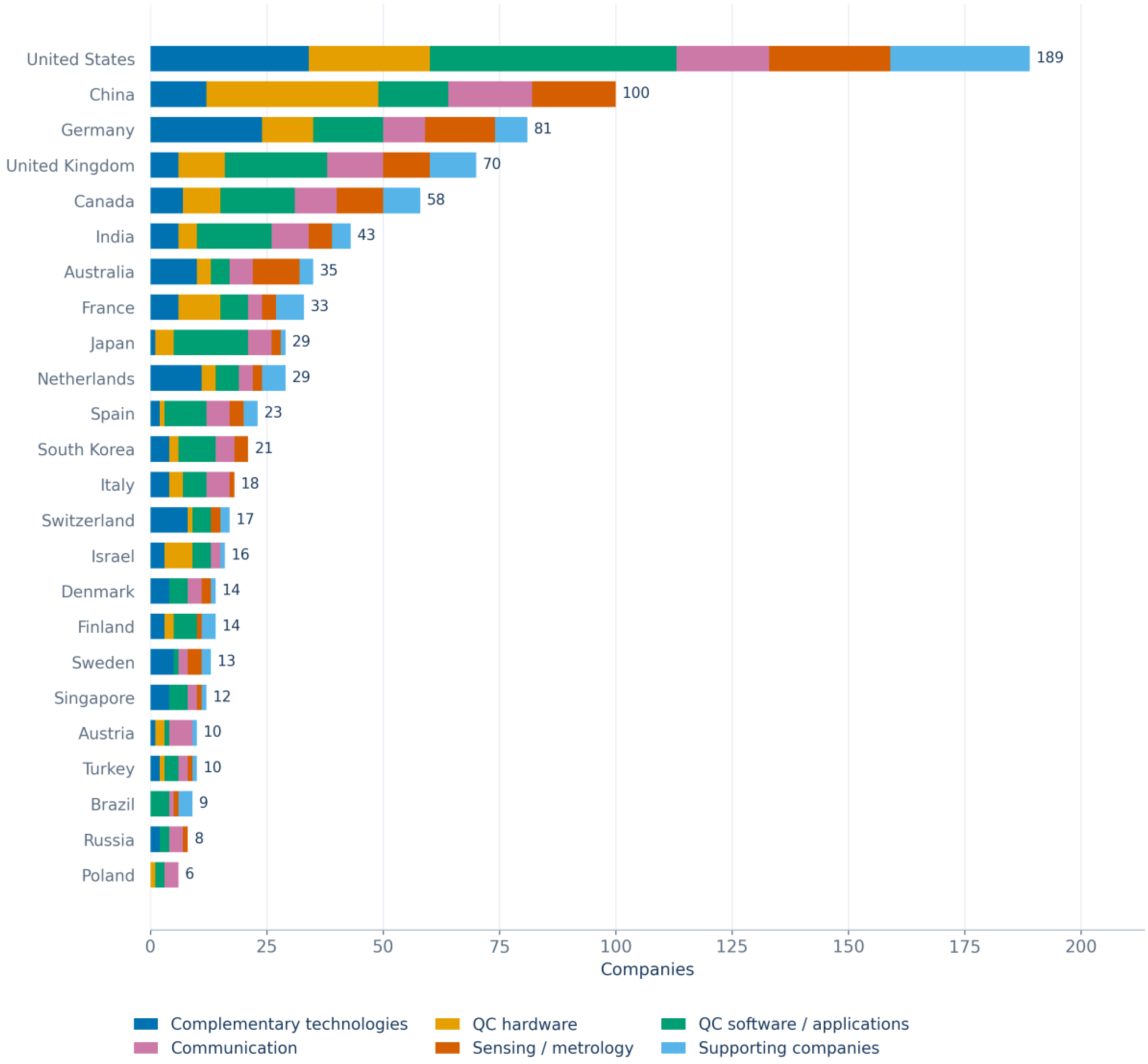

**Figure 8.** Main technology fields in the 24 countries with at least five active QT-origin companies (858 of the full population’s 935 companies). Each company contributes one main field. QT hybrids with a PQC main label are grouped under communication; the separate PQC cohort is excluded.

Regional aggregation puts the EU-27 ahead of the United States, with 263 active companies (28.1%) compared with 189 (20.2%) (Figure 9). The rest of the world is the largest group, with 320 companies (34.2%) across 45 countries and jurisdictions. It was the smallest in 2022, with 87 companies (19.7%) across 19 countries. Expanded geographical coverage contributes to this change alongside more recent company formation.

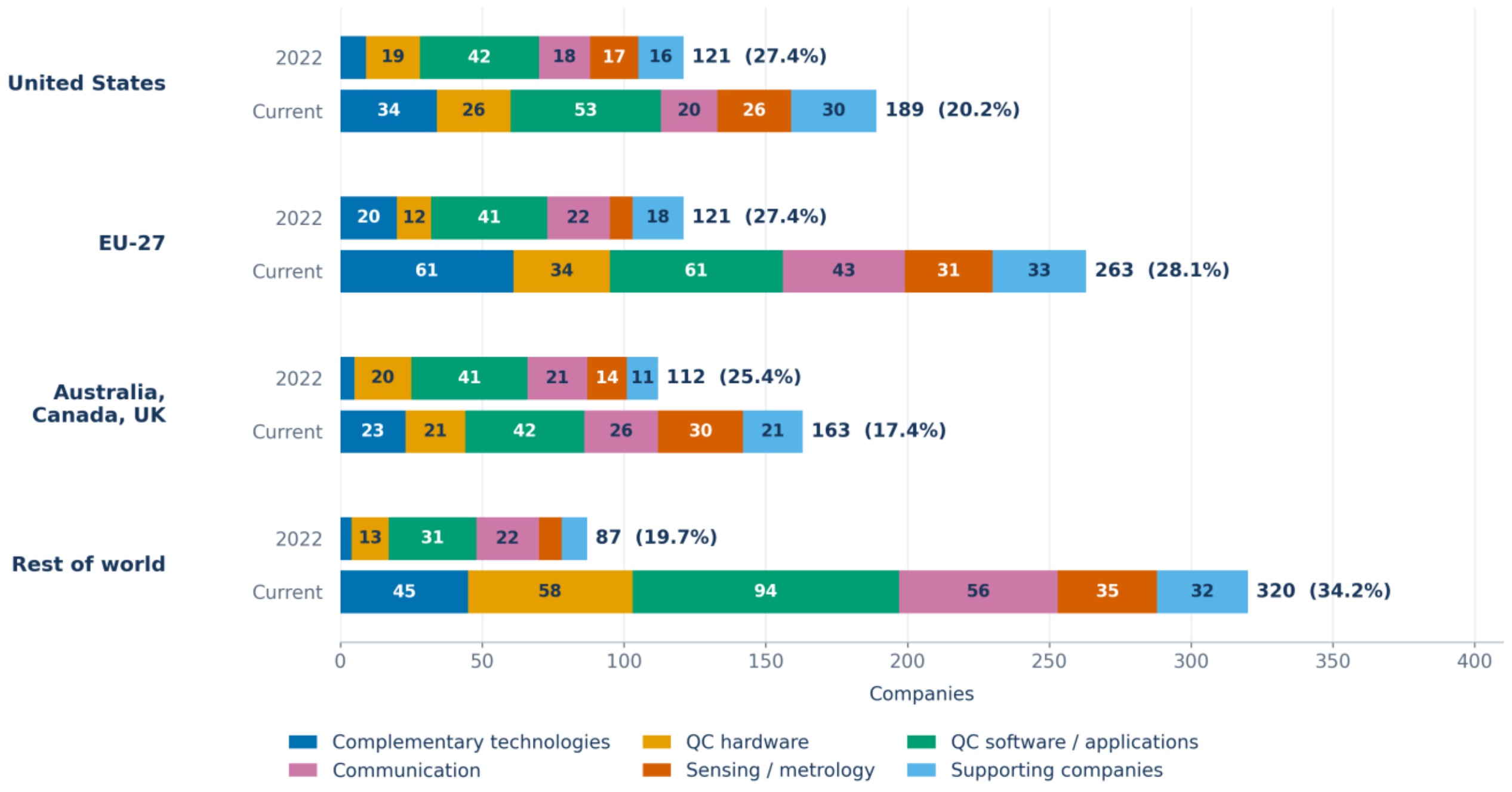

**Figure 9.** Regional technology fields in the 2022 study [1] (n = 441) and all active QT-origin companies in the update (n = 935). Labels give counts and shares of each study total. The earlier population included acquired and inactive companies and PQC; the updated population separates PQC. Discovery coverage also differs.

Software and applications form the largest field in the United States, with 53 companies (Figure 10). In the EU-27, complementary technologies and software share first place, with 61 companies each. China's 100 companies are concentrated in hardware (37), communication (18) and sensing (18), which together account for 73.0% of its population.

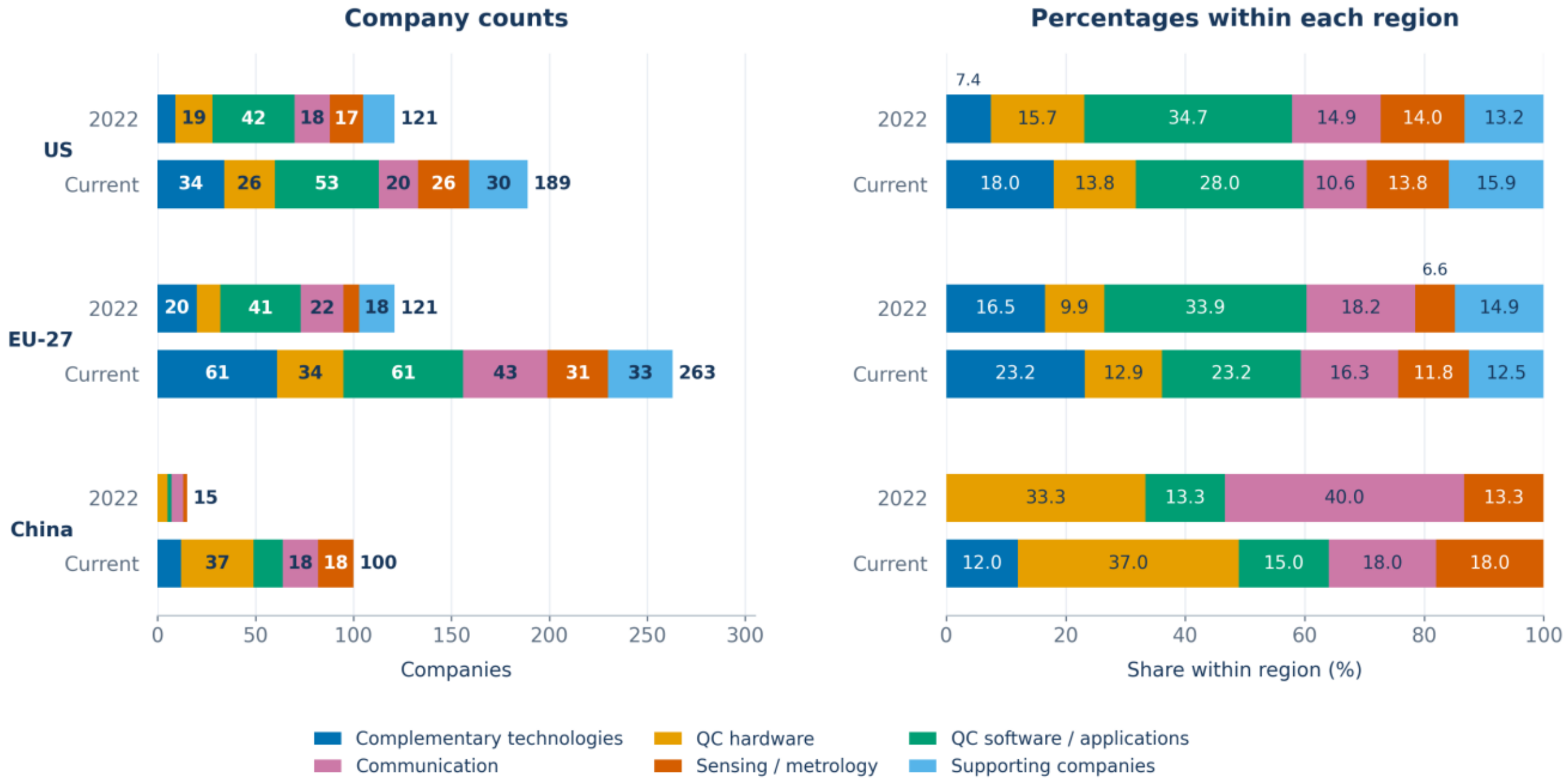

**Figure 10.** Technology fields in the United States, EU-27 and China. Paired bars compare the 2022 study [1] with all active QT-origin companies: 121 and 189 in the United States, 121 and 263 in the EU-27, and 15 and 100 in China. Panels show counts and within-region shares. Coverage and inclusion differences are as in Figure 9.

## 3.3 Technology fields and routes into QT

Software and applications remain the largest main field, accounting for 250 of the 935 active companies (26.7%) (Figure 11). Complementary technologies contribute 163 companies and supporting services 116. Together, these suppliers and service businesses represent 29.8% of the active population.

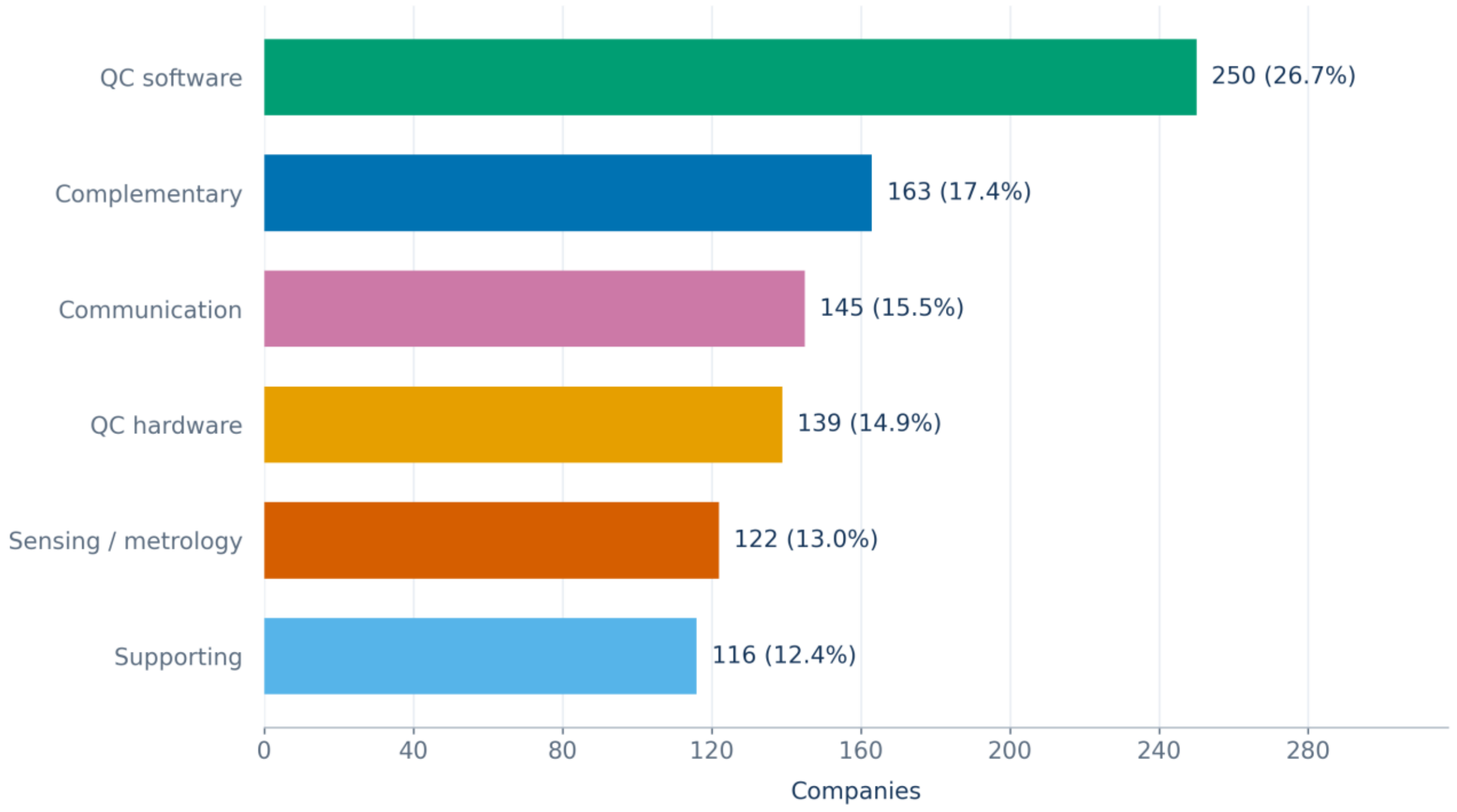

**Figure 11.** Main technology fields of all 935 active QT-origin companies. Each company contributes one field; the separate PQC cohort is excluded.

The separate PQC cohort contains 239 companies, of which 216 are active, eleven acquired, nine possibly inactive and one dissolved; two have unresolved lifecycle status (Figure 12). The United States accounts for 59 companies, followed by Canada and the United Kingdom with 19 each. These counts describe the PQC businesses identified during the broader QT search.

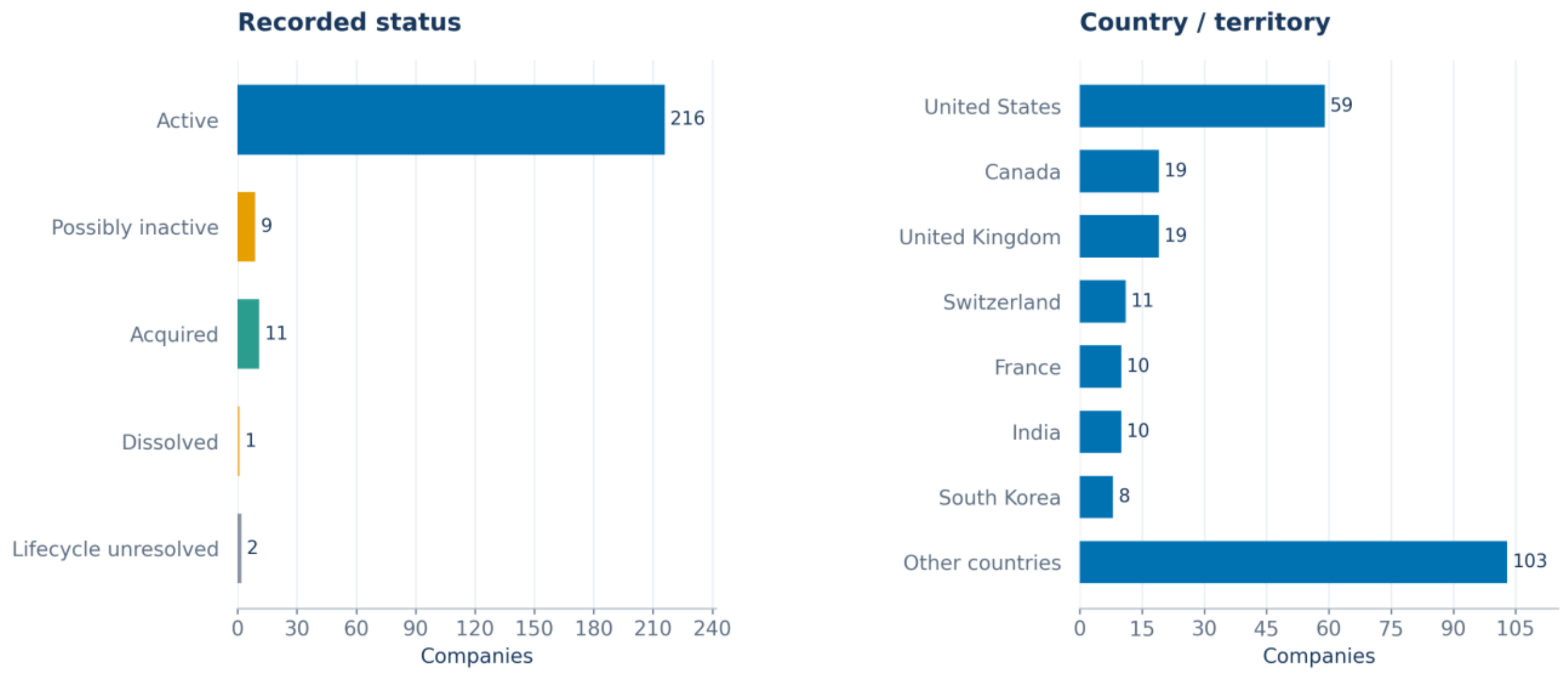

**Figure 12.** Company status and country distribution in the separate PQC cohort (n = 239). All founding years are included. Two lifecycle statuses are unresolved; a pending acquisition is retained as active.

Complementary technologies dominate the later-entrant population, accounting for 253 of its 446 companies (56.7%) (Figure 13). Computing software is the next largest field, with 70 companies.

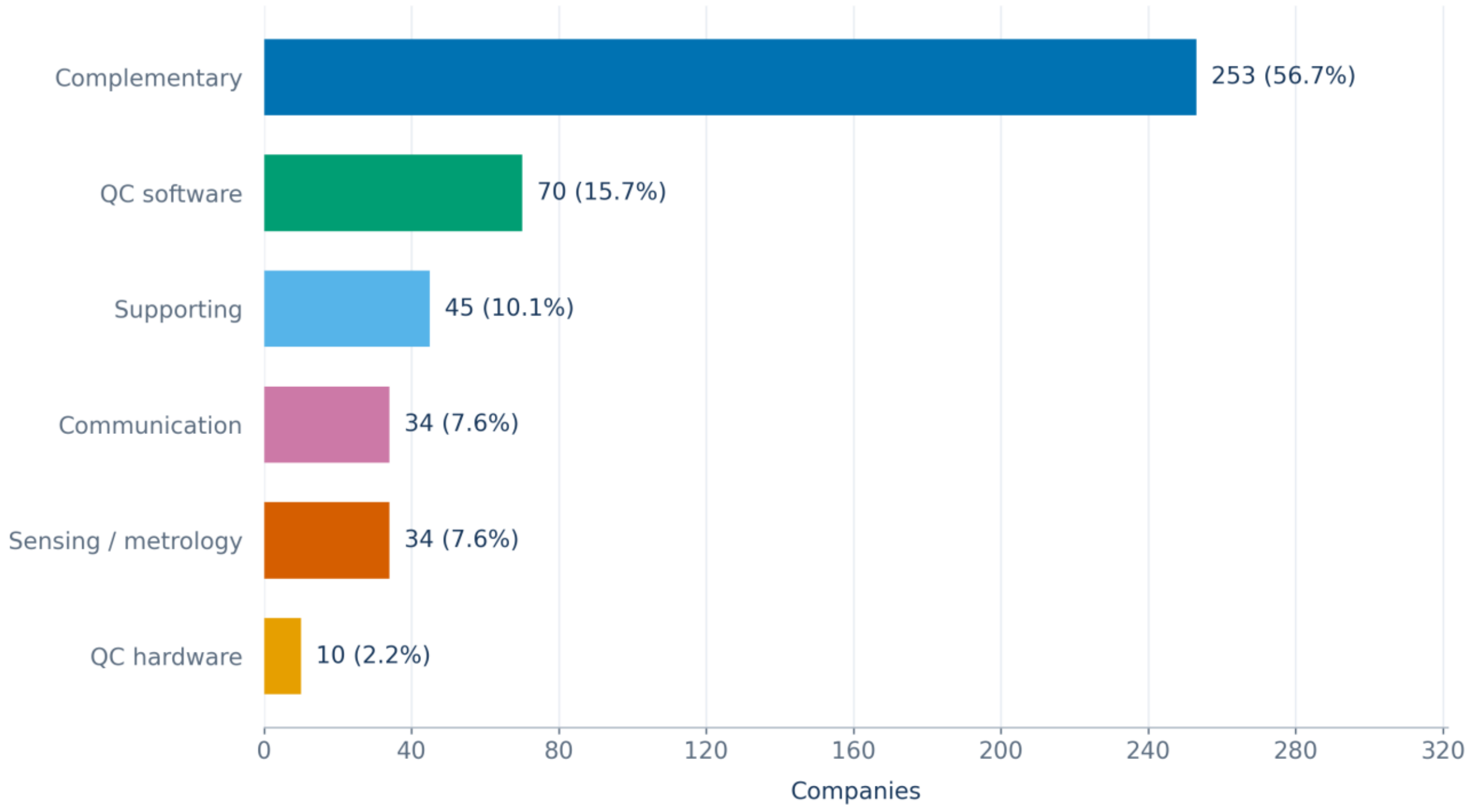

**Figure 13.** Main QT field of 446 later entrants. All founding years and company statuses are included.

Among the three core technical areas, sensing has the highest share of later entrants: 34 of 166 companies (20.5%), compared with 14.1% for computing and 16.7% for communication (Figure 14).

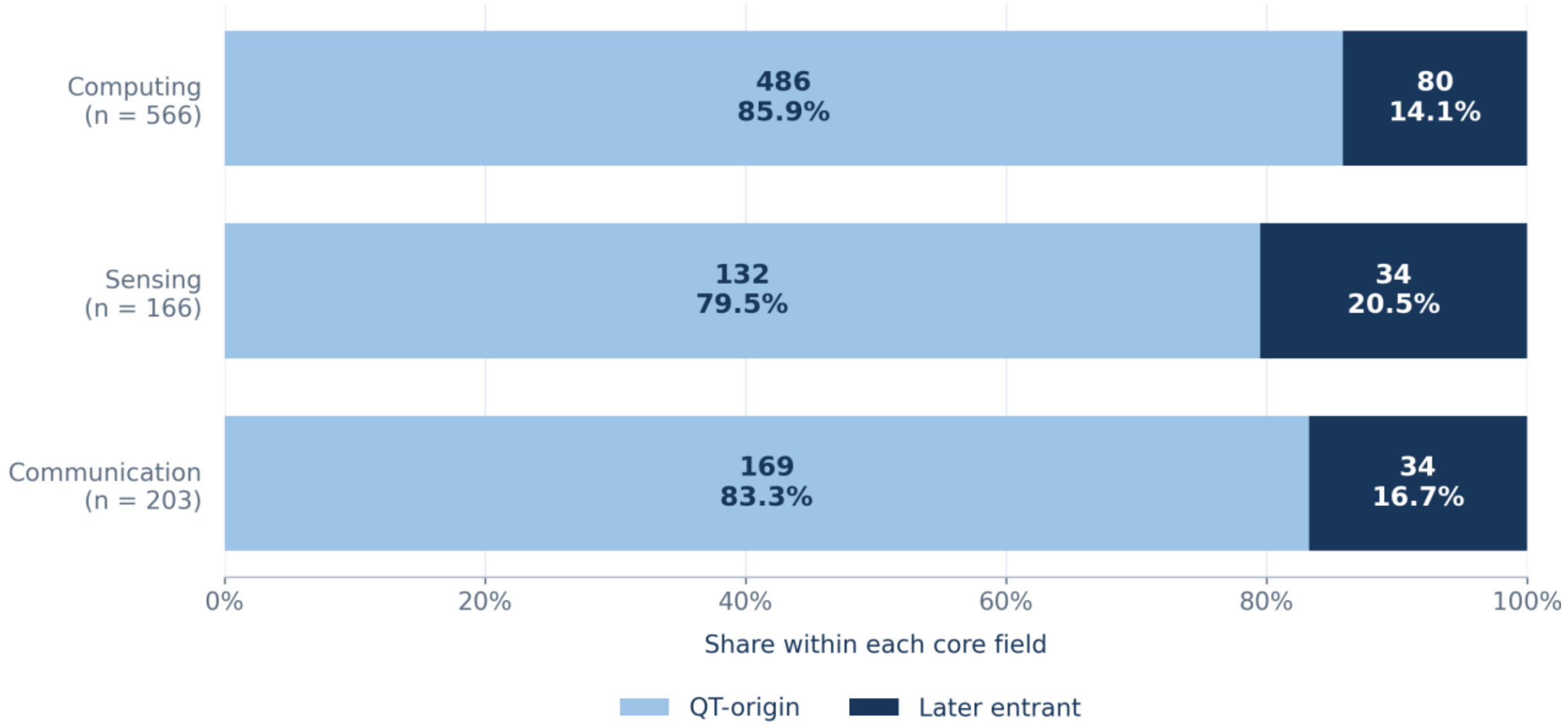

**Figure 14.** QT-origin companies and later entrants across computing, sensing and communication (n = 935). Computing combines hardware and software. Counts include all founding years and company statuses; the separate PQC cohort is excluded.

Established firms also participate across the QT supply chain (Figure 15). Complementary technologies and computing software are particularly prominent in the incumbent review, while sensing accounts for a smaller share.

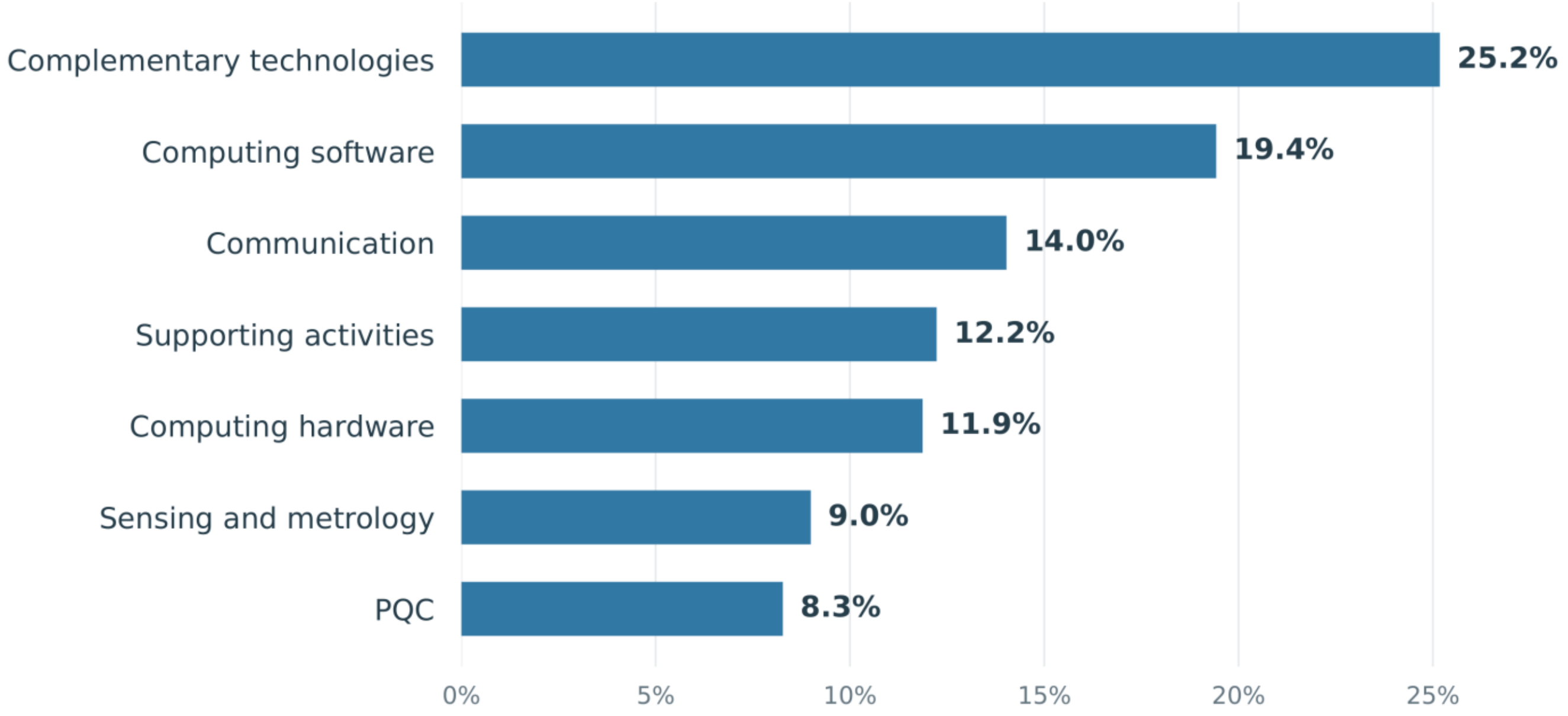

**Figure 15.** Technology fields represented in the incumbent review based on Understanding Quantum Technologies [9]. Each of the 161 company, brand and business-unit entries is counted once in every field in which it participates. Percentages use the resulting 278 counts as the denominator.

## 3.4 Companies and national policy milestones

In 26 of the 28 countries included in the policy comparison, a QT-origin company predates the selected policy milestone (Figure 16). For Slovakia and Luxembourg, the analysed cohort contains no qualifying company with a known founding year through 2025.

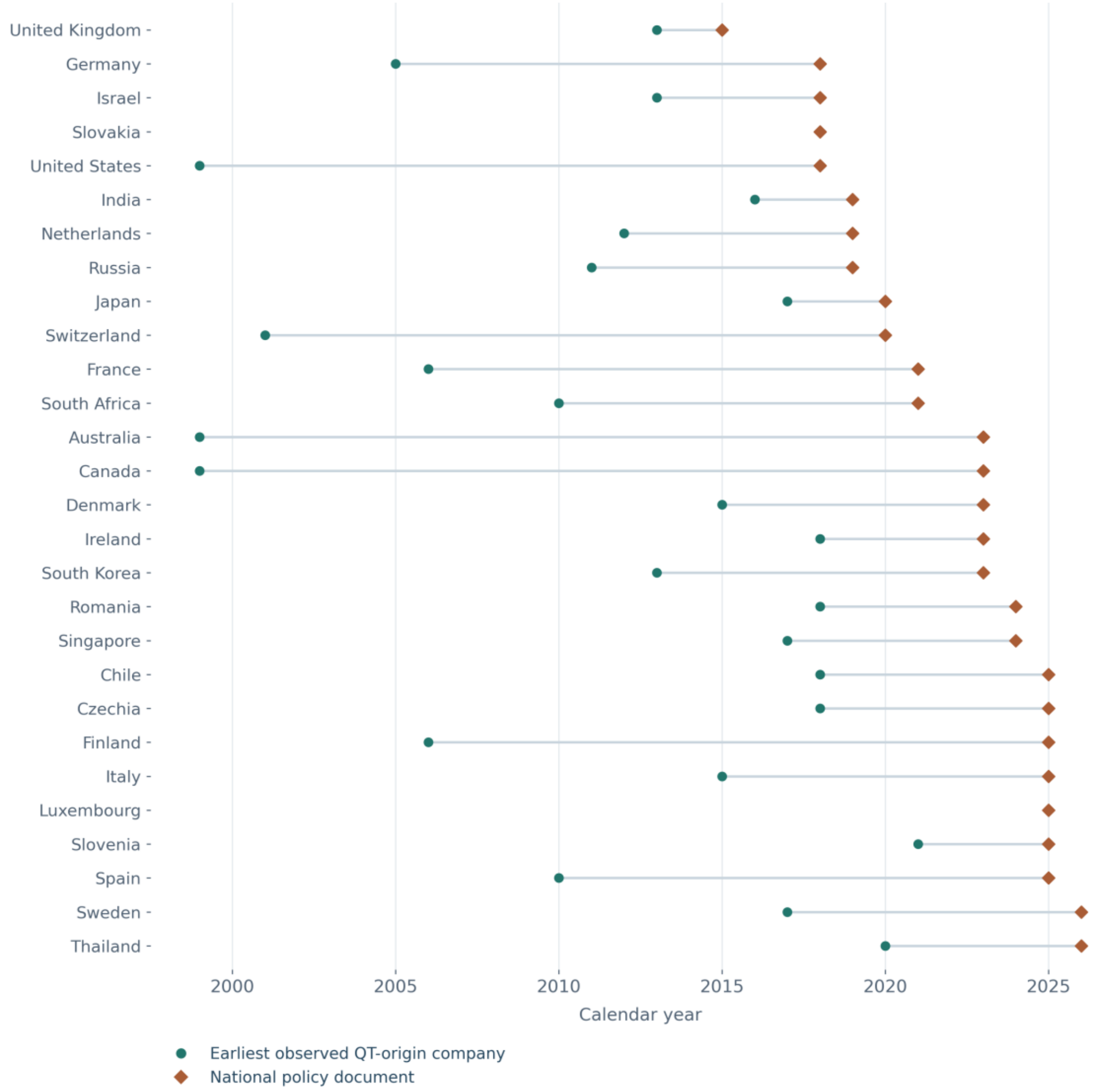

**Figure 16.** Earliest recorded QT-origin company and selected national policy milestone in 28 countries. Company foundations through 2025 include active, acquired or merged, and inactive companies. Missing company points in Slovakia and Luxembourg indicate no qualifying company with a known founding year in this cohort.

Across the twelve countries with complete three-year comparisons, 72 companies were founded before the selected milestones and 165 afterwards (Figure 17). Counts rose in eight countries, fell in two and were unchanged in two. The largest increases were in the United States, from 26 to 57; Germany, from six to 27; and the United Kingdom, from four to 22. Technology-field breadth also generally increased.

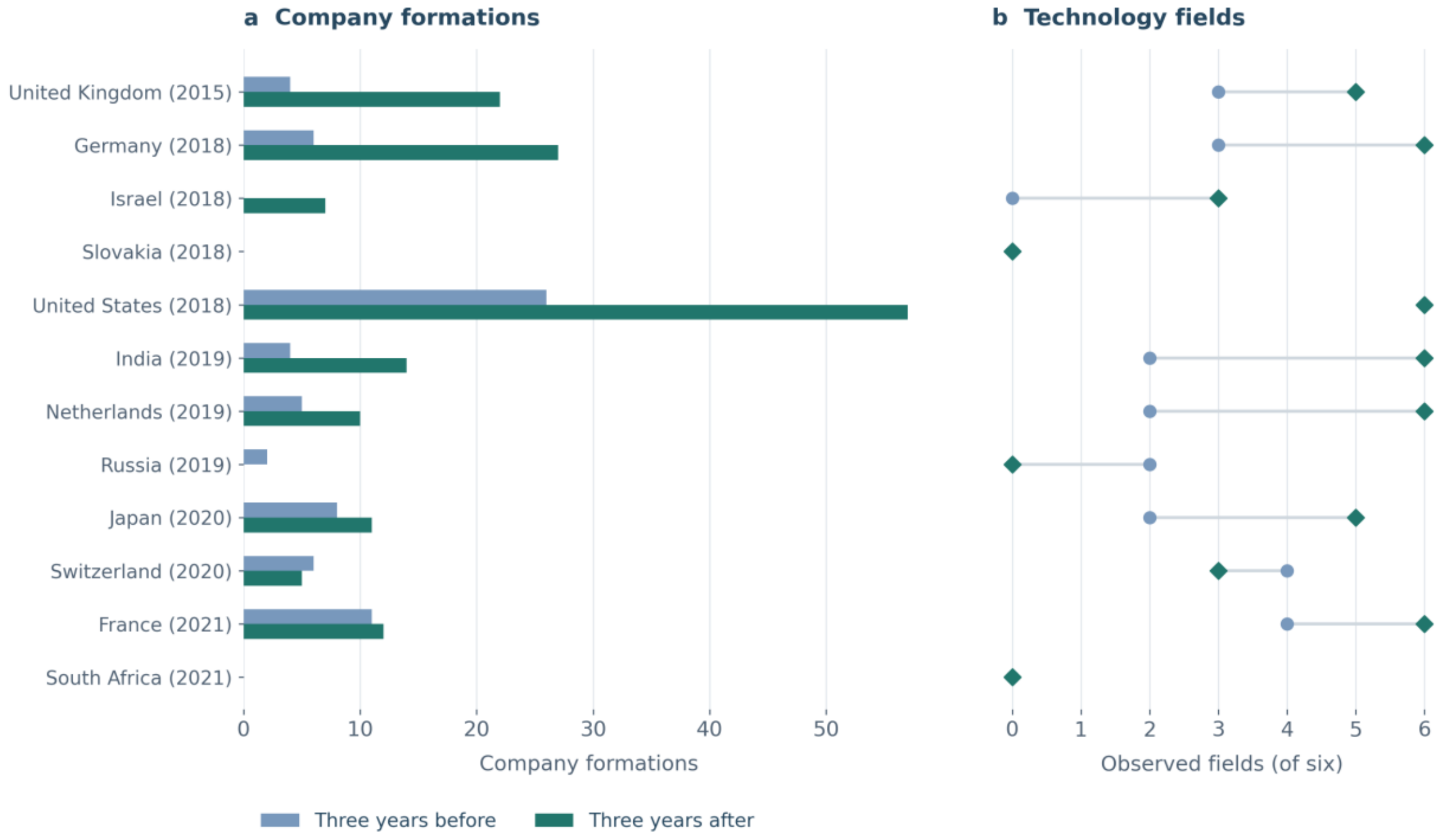

**Figure 17.** Company foundations and represented fields in the three-year periods before and after selected national policy milestones, excluding the milestone year. Twelve countries have complete periods through 2025, with 72 foundations before and 165 after. All lifecycle statuses are included.

The national increases occurred during wider global expansion. Four countries gained a share of worldwide company foundations, six lost share and two were unchanged (Figure S3). Appendix A2 reports the additional statistical comparisons.

## 3.5 Supporting businesses usually followed technical companies

Most supporting businesses followed earlier technical QT companies in their jurisdiction: 101 of 124, or 81.5% (Figure 18). Among the 25 jurisdictions containing both groups, technical companies appeared first in 20, supporting companies in four, and both in the same year in one.

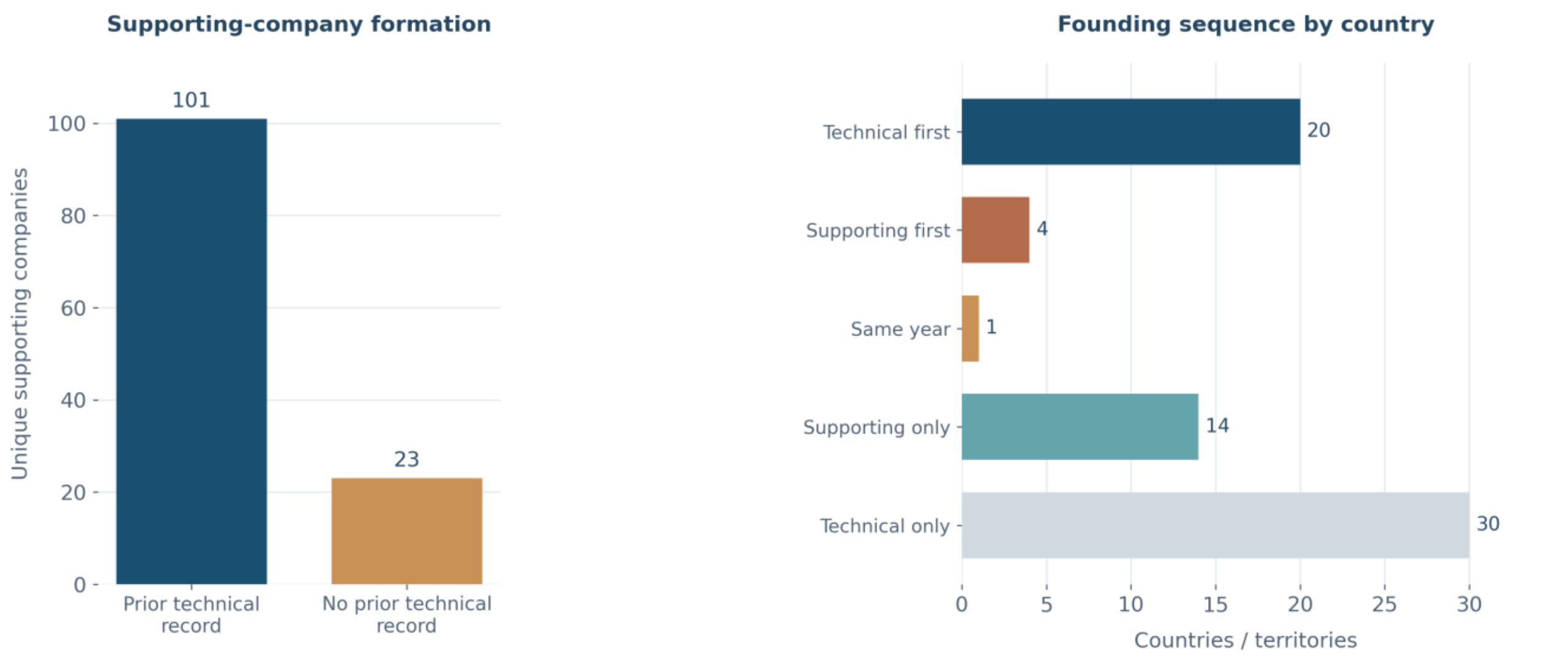

**Figure 18.** Supporting-company foundations relative to technical-company foundations in the same jurisdiction. The cohort contains 124 supporting and 898 technical companies founded through 2025. Prior technical activity requires an earlier founding year. The country comparison covers 69 jurisdictions.

Annual recorded foundations reached 24 supporting businesses and 122 technical companies in 2025 (Figure S4).

## 3.6 AI, defence and space in company activities

Among the 1,092 QT-origin companies, documented AI relevance rises from 344 (31.5%) in the earliest assessed evidence to 483 (44.2%) in their current or last recorded activities (Figure 19).

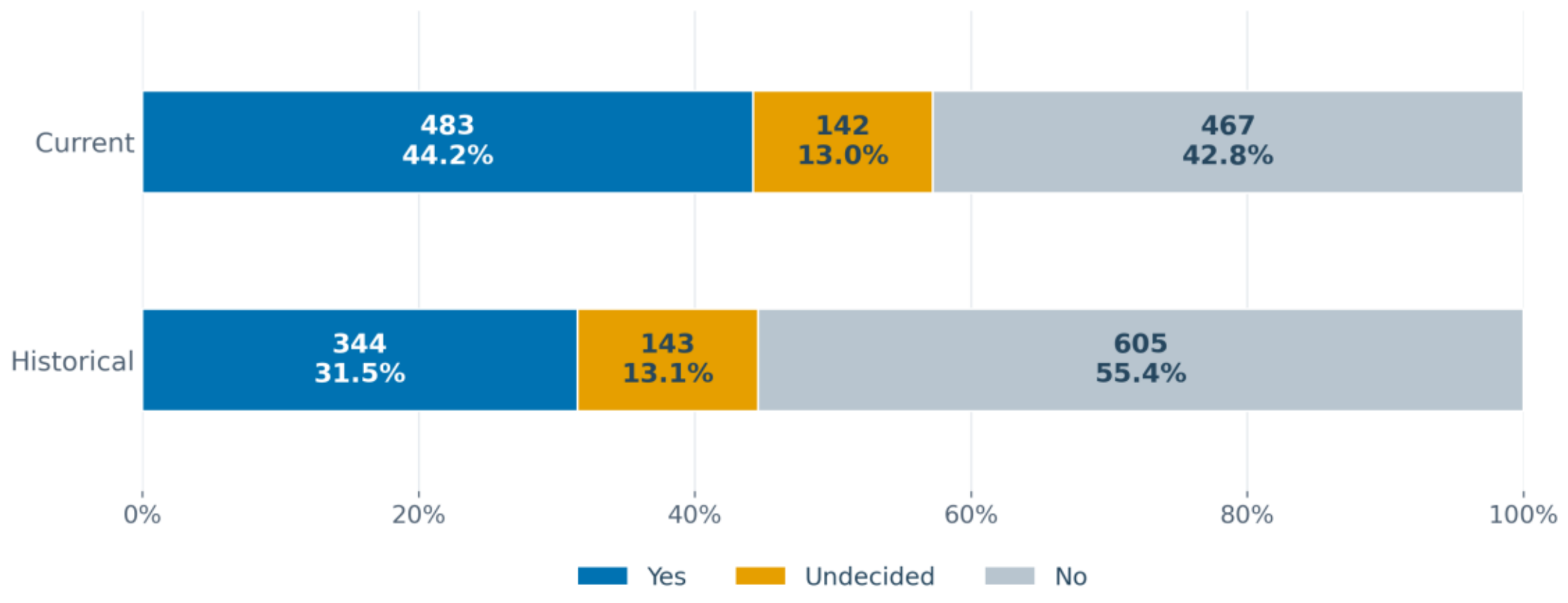

**Figure 19.** Historical and current AI relevance among all 1,092 QT-origin companies across lifecycle statuses. Historical means earliest assessed activity; current means present or last recorded activity.

Defence involvement rises from 87 of the 1,092 companies (8.0%) historically to 417 (38.2%) currently (Figure 20). This increase of 30.2 percentage points exceeds the 12.7-point increase for AI, although AI remains more common in current activities. Among active companies, 390 have current defence relevance across 42 countries and jurisdictions (Figure S5).

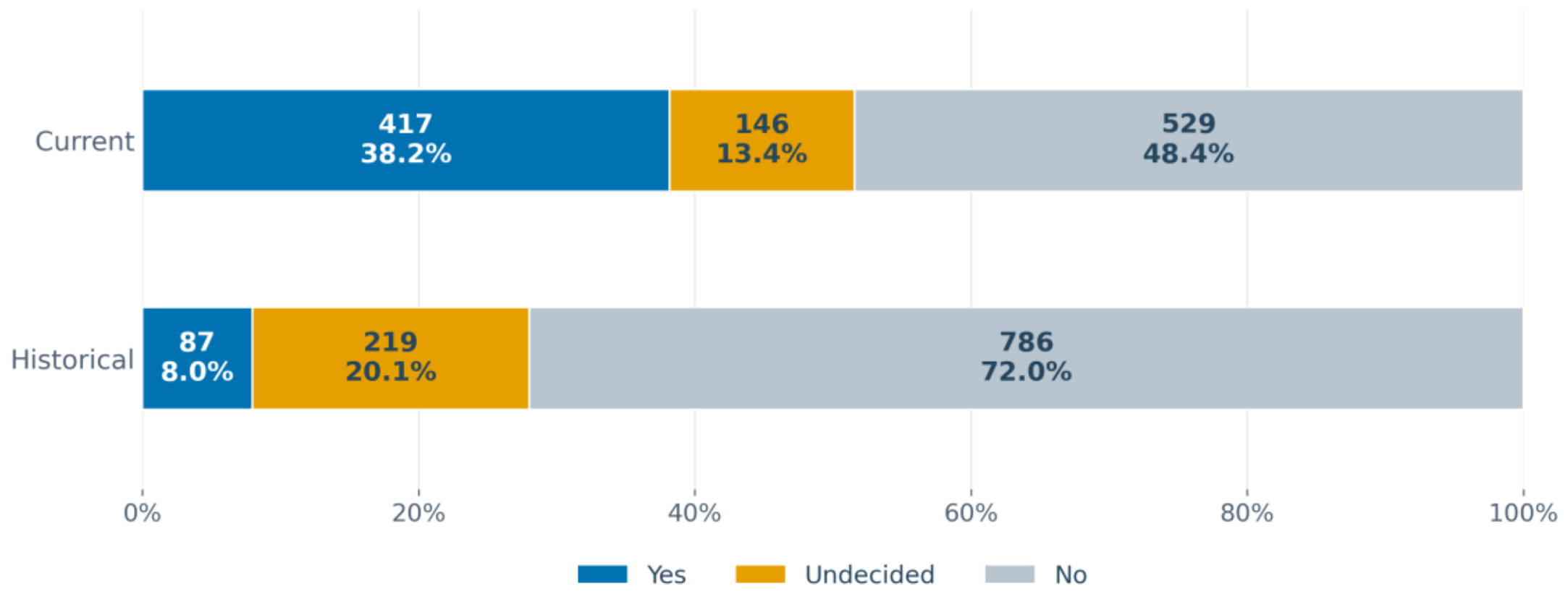

**Figure 20.** Historical and current defence relevance among QT-origin companies across all lifecycle statuses (n = 1,092). Historical means earliest assessed activity; current means present or last recorded activity.

Documented space relevance rises from 17.4% historically to 28.3% currently among the active and acquired companies reviewed (Figure 21). Among active companies, 266 (28.4%) have space relevance across 36 countries and jurisdictions. It is most common in communication (46.2%) and sensing (41.8%). Most space-relevant active companies also have defence relevance (195 of 266; 73.3%).

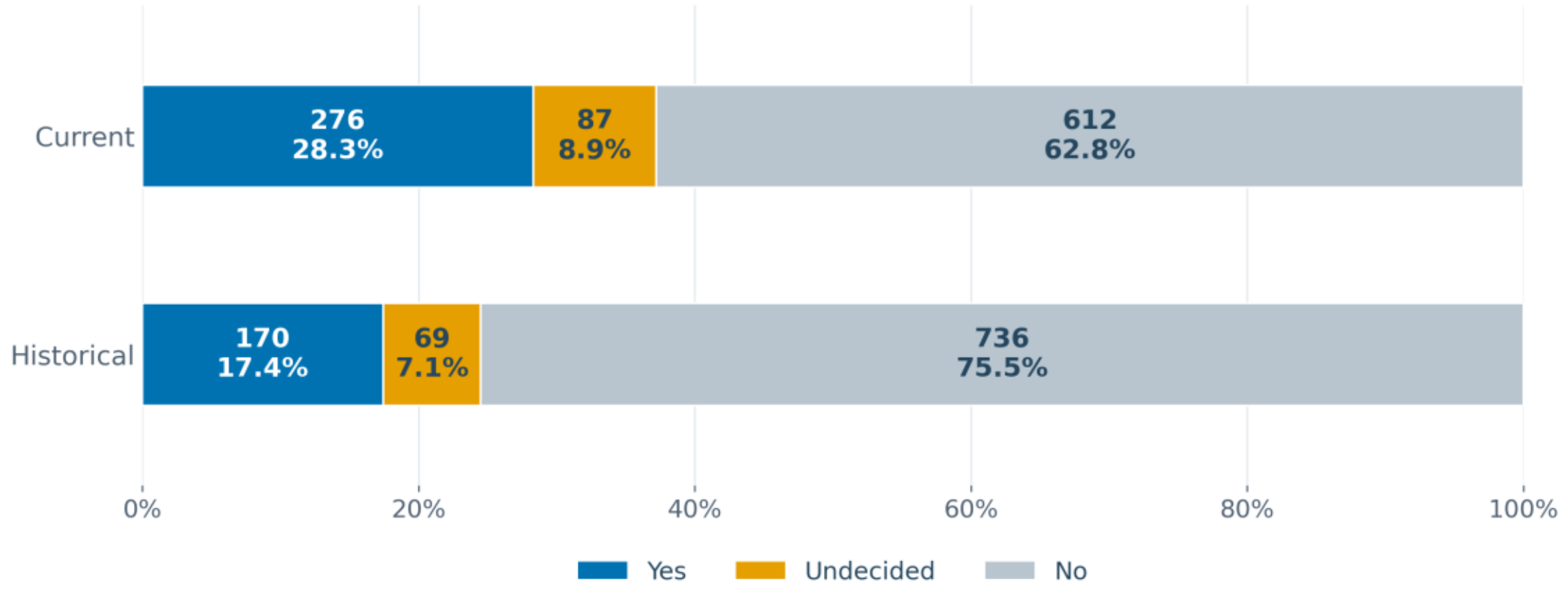

**Figure 21.** Historical and current space relevance among 975 active or acquired/merged QT-origin companies. The 117 inactive QT-origin companies have no space classifications and are excluded. Historical means earliest assessed activity; current means present or last recorded activity.

Among active companies founded through 2025, 265 of the 380 currently defence-relevant companies (69.7%) have a recorded pivot into defence (Figure 22). The share is lower for AI, at 111 of 400 companies (27.8%), and for space, at 114 of 259 (44.0%).

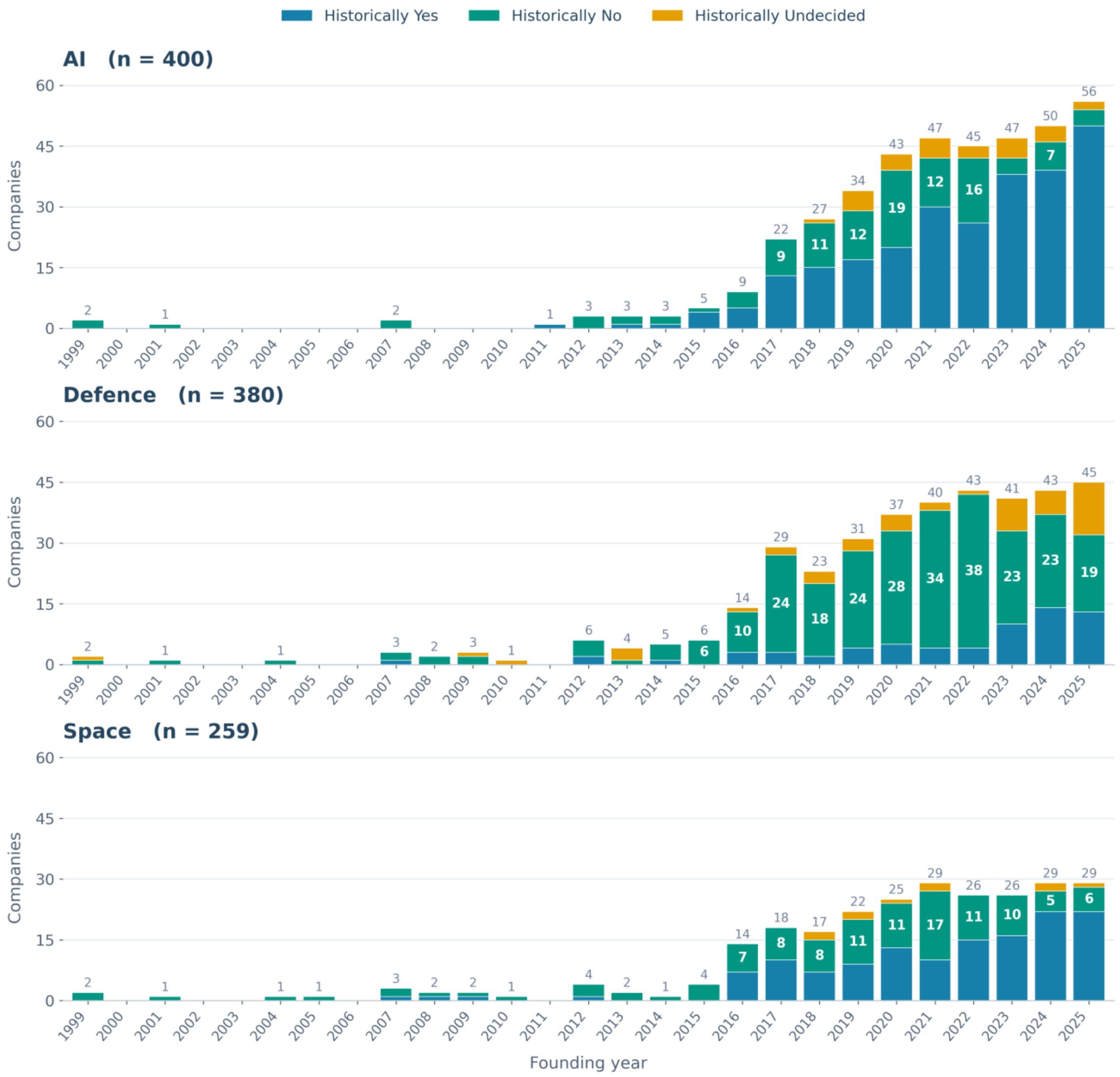

**Figure 22.** Historical classifications of currently AI-relevant (n = 400), defence-relevant (n = 380) and space-relevant (n = 259) active QT-origin companies founded through 2025. Historical No followed by current Yes is a recorded pivot. Undecided combines historical Possibly and Uncertain. Founding year does not date the transition.

AI and defence pivots overlap only partly. Among active companies with known founding years, 58 have an AI pivot without a defence pivot, 214 a defence pivot without an AI pivot, and 54 both (Figure 23). Recorded defence pivots are particularly common in the 2015-2019 cohort, at 43.4% including dual pivots, compared with 17.7% among companies founded in 2023-2025. Older companies have had longer to change and longer histories in which to document those changes.

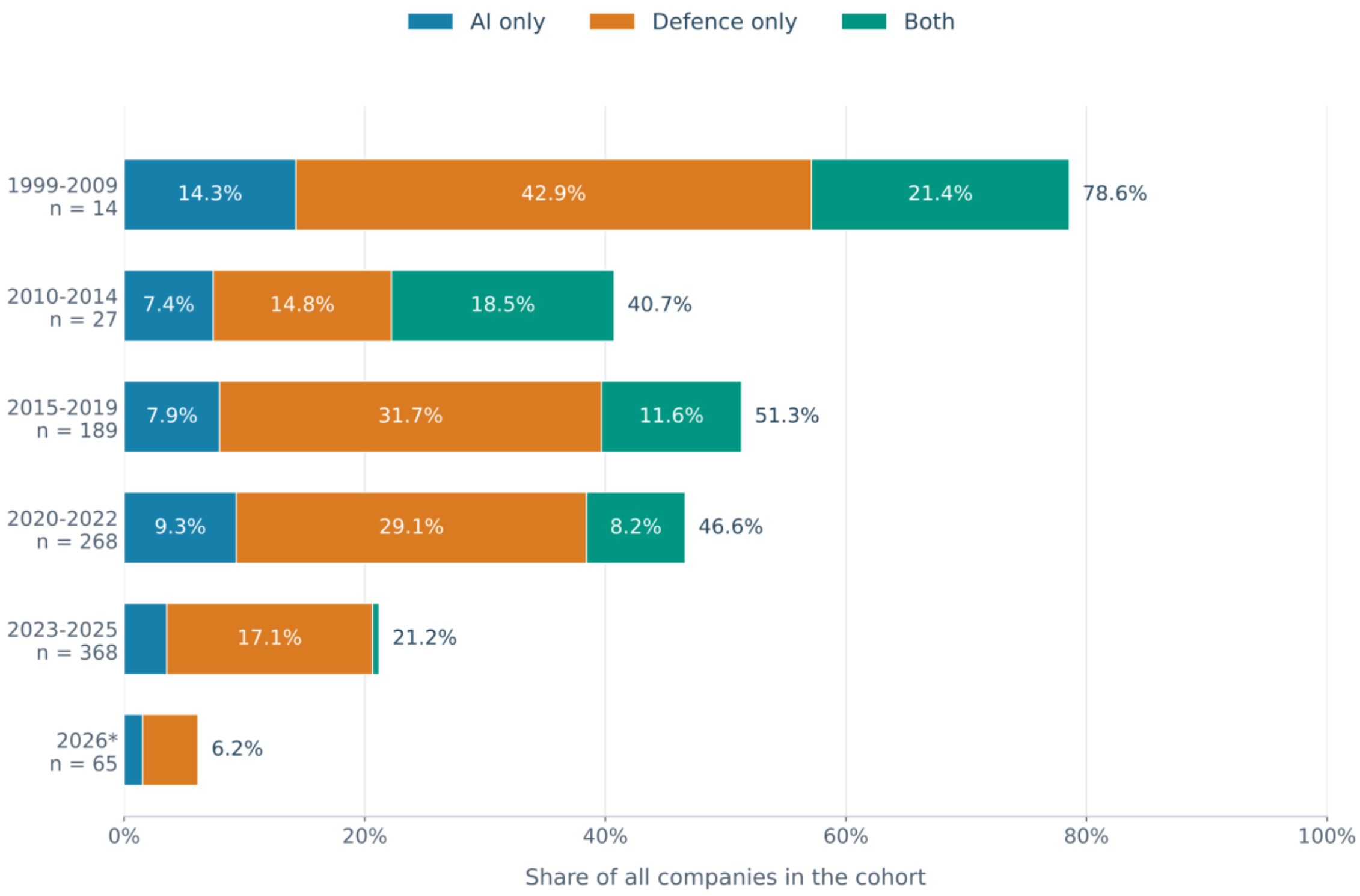

**Figure 23.** Recorded pivots into AI only, defence only or both, as shares of all active QT-origin companies in each founding cohort (n = 931). Categories are mutually exclusive within the AI and defence comparison, irrespective of space relevance. Undecided cases remain in cohort denominators. The asterisk marks the incomplete 2026 cohort. Founding years do not date pivots.

## 3.7 Public listings and financial scale

Eleven public companies meet the historical and current QT criteria and exceed US$500 million in market capitalisation in the 16 October 2026 comparison (Figure 24). Nine are listed on North American exchanges and two in Shanghai. These listings include companies from other regions: IQM and Pasqal, for example, are European companies.

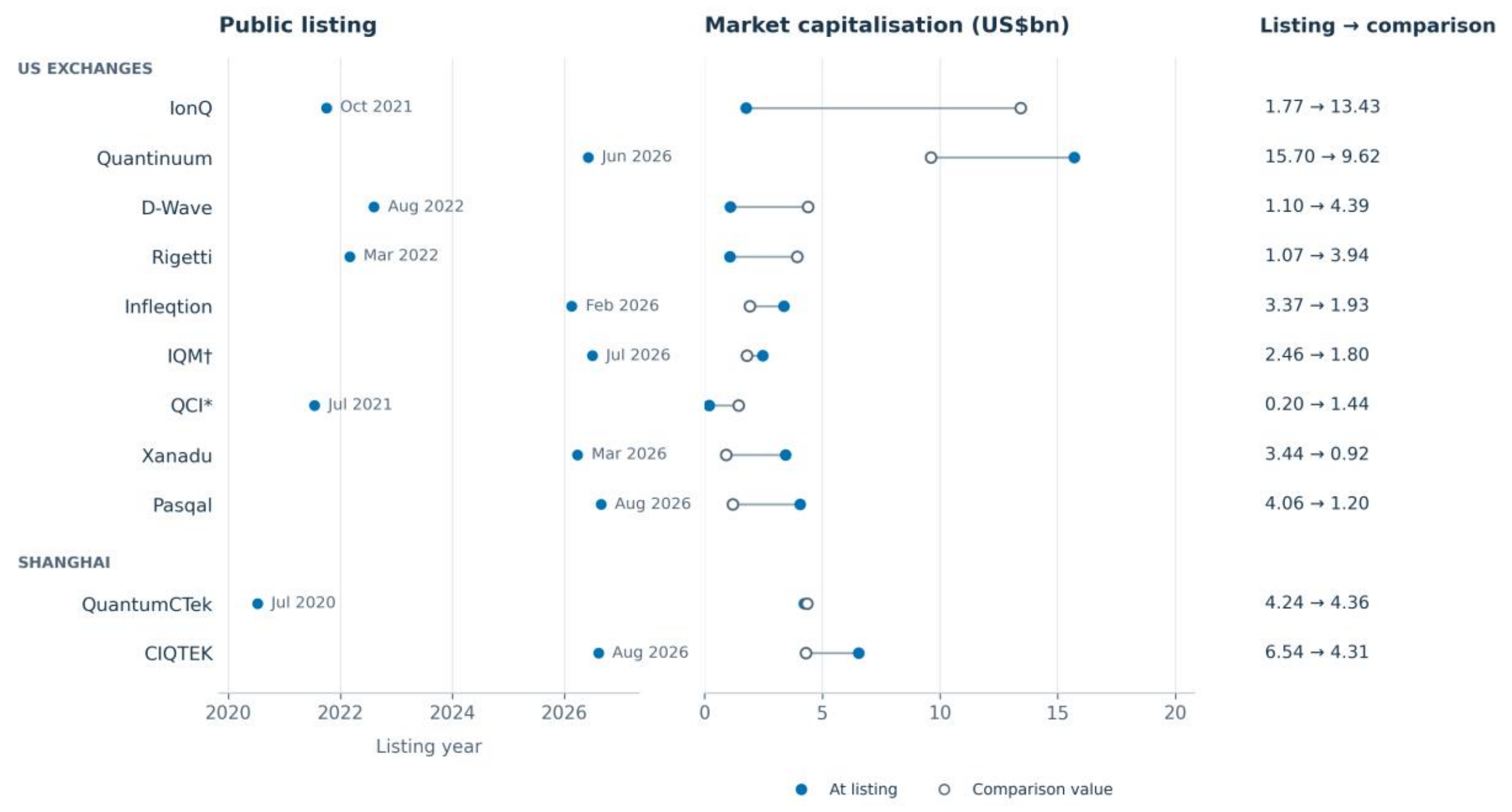

**Figure 24.** Listing and 16 October 2026 market capitalisation for eleven selected public QT companies above US$500 million. Filled circles indicate listing and open circles the comparison value at the regular trading close. Baselines are first trading-day closes; * marks QCI's Nasdaq uplisting and † IQM's additional Helsinki listing. Source details are in Appendix A4.

IonQ has the largest reported market capitalisation in this selection, at approximately US$13.43 billion, compared with US$1.77 billion at its 2021 debut. Quantinuum follows at US$9.62 billion. IonQ alone therefore exceeds the combined US$3.00 billion of the two European-origin companies in the figure, IQM and Pasqal. Six of the eleven companies began trading in 2026, and all six were below their first-day closing market capitalisation at the comparison date.

**Table 2.** Selected private QT-origin companies with reported financing valuations above US$500 million.

| Company | US$ billion | Financing date | Valuation basis and source |
|---|---|---|---|
| PsiQuantum | 7.00 | Sep 2025 | Financing value; company announcement [10] |
| SandboxAQ | 5.75 | Apr 2025 | Reported financing value [11] |
| Origin Quantum | 3.13 | Jun 2026 | Pre-money, CNY21bn; investor announcement [12] |
| Photonic | 2.00 | May 2026 | Post-money; company announcement [13] |
| TuringQ | >1.04 | 2026 (reported) | Reported lower bound, >CNY7bn [14] |
| QuEra | 0.75–1.00 | Feb 2025 | Reported financing range; unconfirmed by company [15] |
| Quantum Machines | 0.70 | Feb 2025 | Reported financing estimate [16] |
| Multiverse Computing | >0.50 | Jun 2025 | Reported lower bound for completed Series B; excludes the subsequently announced open Series C [17] |

Values are approximate. Pre-money excludes the new capital; post-money includes it. CNY values use the 6 October 2026 exchange rate [18]. The table covers completed financing events reported by that date.

The largest private financing valuations identified are US$7.00 billion for PsiQuantum [10] and US$5.75 billion for SandboxAQ [11] (Table 2). Intel agreed in October 2026 to acquire Classiq for approximately

US$1.9 billion, with up to US$350 million in additional payments tied to performance milestones. The transaction is expected to close in the first half of 2027 [19].

# 4 Discussion

## 4.1 What the updated landscape shows

**RQ1: What are the temporal and geographical distributions of QT companies?**

We now identify QT companies in more countries, and recent founding cohorts are substantially larger than the apparent 2018 peak in the earlier study. The update captures both new company formation and older companies missed by the earlier search.

**RQ2: How can QT companies be categorised, and how are these categories populated?**

The six fields continue to capture the main activities of QT companies. Software and applications form the largest group, while suppliers and supporting businesses together account for almost a third of active companies. Most later entrants operate in complementary technologies. The separate PQC population also shows the commercial activity surrounding quantum-resistant classical cryptography.

**RQ3: What patterns can be derived from empirical data on trends?**

Companies usually appeared before the selected national policy milestones, and supporting businesses usually followed domestic technical companies. AI, defence and space are more prominent in companies' current documented activities than in the earliest material we could assess.

## 4.2 Revisiting the arguments about ecosystem development

**A1: Early company activity may precede or catalyse national initiatives, followed by more companies and greater diversity of technology fields.**

Companies preceded most selected policy milestones, and formation was generally higher afterwards. This occurred during a broader global expansion: several countries added companies while their share of worldwide foundations fell.

**A2: Established firms may limit opportunities for new quantum-sensing companies.**

Later entrants account for a greater share of sensing companies than of computing or communication companies. Established firms also participate in sensing, as the incumbent review shows. Whether their presence limits opportunities for new companies remains an open question.

**A3: Growing and diversifying industry needs support the emergence of service companies.**

Supporting businesses generally appeared after technical companies in the same jurisdiction, consistent with the sequence envisaged in 2022. In some jurisdictions, however, they were the first or only QT companies identified, showing that service provision can also be an early form of commercial participation.

**A4: Regional comparisons reveal patterns that country rankings obscure.**

The United States leads the country ranking, while the EU-27 contains more companies when treated as a region, and the combined rest of the world is larger still. The EU-27 also combines distinct national profiles:

complementary suppliers form Germany’s largest field, while France’s smaller population is more evenly spread across fields (Figure 8). China’s concentration in hardware, communication and sensing differs from the larger software and supplier populations in the United States and EU-27.

**A5: Classification may require greater granularity as the ecosystem develops.**

We have separated PQC, as anticipated in 2022. The six fields still provide a useful overview of QT companies, while company histories and involvement in AI, defence and space add further detail. Studies of particular products or technological approaches may need finer categories.

**A6: Company counts alone cannot represent economic significance.**

IonQ’s market capitalisation exceeds that of several other listed QT companies combined, while some private companies have reached multibillion-dollar financing values. These differences in financial scale matter when assessing what country or regional company totals represent.

**A7: Nonprofits, consortia and other supporting organisations warrant their own ecosystem study.**

Nonprofits, consortia, research infrastructures and educational organisations are part of the communities around commercial QT [8]. Grassroots quantum education combines skills development with commitments to access, inclusion and diversity [20]. UNESCO’s account of the International Year of Quantum Science and Technology places participation in a wider international context [21]. The European Quantum Academy, launched in 2026 to coordinate quantum education and professional upskilling, provides a recent example of workforce development [22]. A broader ecosystem study could examine how these organisations’ work supports commercial QT.

## 4.3 Diversification and consolidation

**PQ1: Has diversification continued, or has consolidation become dominant?**

New companies continue to appear across all six fields while others merge, are acquired, close or leave QT. Diversification and consolidation are therefore happening at the same time.

The historical comparisons distinguish AI activity already present in many companies’ earliest records from defence activity more often documented later. Space relevance is particularly common in communication and sensing and often overlaps with defence. The defence findings sit alongside publicly stated interest in military quantum navigation, sensing, communication and computing [23], NATO’s quantum strategy [24] and its emphasis on cooperation with industry [25].

Fifteen companies in our register have left QT, several of them turning towards AI or other classical digital products. Qubrid advertised quantum-computing services in 2023 [26] and now emphasises graphics processing unit (GPU) infrastructure and AI model services [27]. Icosa Computing, which began by developing quantum annealing algorithms, now develops local AI tools drawing on classical optimisation methods [28].

**PQ2: Was 2018 a lasting peak in company formation?**

Company formation continued well beyond 2018. The larger recent cohorts extend across all six fields, with 2025 the largest complete-year cohort.

## 5 Conclusions

Commercial QT now involves developers, suppliers and service businesses following different routes into and through the industry. We identify 935 active QT-origin companies across 72 countries and jurisdictions within a wider register of 1,792 eligible companies. Company formation continued beyond the apparent 2018 peak, with recent cohorts larger than those observed in the earlier study. New companies continue to appear across all six technology fields alongside acquisitions, closures and changes of focus. Diversification and consolidation are both part of this development.

Distinguishing companies by origin and activity reveals different forms of commercial participation. Later entrants are concentrated in complementary technologies, while suppliers and supporting businesses together account for almost a third of the active population. AI activity was already present in many companies' earliest records, whereas defence involvement more often appeared later. Space activity is particularly common in communication and sensing and frequently overlaps with defence. These histories show commercial QT developing through connections with other fields as well as through new company formation.

The wider geographical coverage also brings different national and regional profiles into view. The United States leads the country ranking, while the EU-27 contains more companies in aggregate. Comparing these counts with technological composition and company valuations reveals differences in specialisation and financial scale that a ranking by company numbers alone would obscure.

## Declarations

### Availability of data and materials

The dataset supporting this study is available from the corresponding author on reasonable request.

### Competing interests

The authors declare that they have no competing interests.

### Funding

The authors received no specific funding for this study.

### Authors' contributions

ZCS conceptualised the study, collected and analysed the data, prepared the figures, and drafted and revised the manuscript. OK collected data. AUA conceptualised the study and contributed to writing and revising the manuscript.

### Acknowledgements

We thank Tobias Denzler, Philipp Sonnenschein and Elias X. Huber for their constructive feedback and contributions to the dataset. We also thank Olivier Ezratty for (unknowingly) giving us one more reason to reopen the dataset just as we thought it was finished.

# Appendix A Additional methods and policy documents

## A1 Classification validation

We manually assessed 100 randomly sampled Yes classifications and 100 randomly sampled No classifications for each of the defence, AI and space relevance classifiers. This yielded 200 assessments per classifier and 600 classification assessments overall, drawn from the 990-company review population. Each assessed case was weighted by the size of its original label group divided by the number sampled from that group. Table A1 reports the resulting weighted performance estimates.

**Table A1.** Weighted classifier performance based on manual validation. TP and TN denote confirmed positive and negative labels; FP and FN denote incorrect positive and negative labels. All calculations use weighted counts.

| Metric | Calculation | AI | Defence | Space |
|---|---|---|---|---|
| Precision | TP / (TP + FP) | 92% | 97% | 94% |
| Recall | TP / (TP + FN) | 97.99% | 97.64% | 93.35% |
| False-negative rate | FN / (TP + FN) | 2.01% | 2.36% | 6.65% |
| False-positive rate | FP / (FP + TN) | 7.95% | 2.55% | 2.70% |

Possibly and Uncertain labels remained separate from Yes and No and were excluded from this validation. These estimates therefore describe current binary classifications within the review population, rather than performance across all classification categories or historical labels.

## A2 National policy comparisons

Table A2 lists the selected documents and milestone years. Asterisks identify the twelve countries with complete three-year periods before and after the milestone through 2025. We omit the milestone year. For Canada, we use the 2023 strategy launch, although the linked report bears a 2022 imprint [29]. For India, we use the 2019 draft concept note [30].

**Table A2.** Selected national policy documents and milestone years. Asterisks identify the twelve complete three-year comparisons.

| Country | Year | Selected document |
|---|---|---|
| United Kingdom * | 2015 | National Strategy for Quantum Technologies [31] |
| Germany * | 2018 | Quantum Technologies: From Basic Research to Market [32] |
| Israel * | 2018 | Quantum Science and Technology: Steering Advisory Committee Final Report [33] |
| Slovakia * | 2018 | QUTE.SK Platform Action Plan [34] |
| United States * | 2018 | National Strategic Overview for Quantum Information Science [35] |
| India * | 2019 | National Mission on Quantum Technologies and Applications: Draft Concept Note [30] |
| Netherlands * | 2019 | National Agenda for Quantum Technology [36] |

| Country | Year | Selected document |
|---|---|---|
| Russia * | 2019 | Quantum Technologies Roadmap [37] |
| Japan * | 2020 | Quantum Technology and Innovation Strategy [38] |
| Switzerland * | 2020 | White Paper: Quantum Technologies in Switzerland [39] |
| France * | 2021 | National Strategy for Quantum Technologies [40] |
| South Africa * | 2021 | South African Quantum Technology Initiative Framework [41] |
| Australia | 2023 | National Quantum Strategy [42] |
| Canada | 2023 | National Quantum Strategy [29] |
| Denmark | 2023 | Strategy for Quantum Technology, Part 1 [43] |
| Ireland | 2023 | Quantum 2030 [44] |
| South Korea | 2023 | National Quantum Strategy [45] |
| Romania | 2024 | National Strategy in the Field of Quantum Technologies 2024–2029 [46] |
| Singapore | 2024 | National Quantum Strategy: Official Media Factsheet [47] |
| Chile | 2025 | Chile Has Quantum: National Quantum Technologies Strategy [48] |
| Czechia | 2025 | National Strategy for Quantum Technologies 2026–2030 [49] |
| Finland | 2025 | Quantum Technology Strategy 2025–2035 [50] |
| Italy | 2025 | Italian Strategy for Quantum Technologies [51] |
| Luxembourg | 2025 | Luxembourg’s Quantum Strategy [52] |
| Slovenia | 2025 | Strategy for the Development of Quantum Technologies until 2035 [53] |
| Spain | 2025 | Quantum Technologies Strategy 2025–2030 [54] |
| Sweden | 2026 | Sweden’s Strategy for Quantum Technology [55] |
| Thailand | 2026 | National Quantum Economy Strategy (approval announcement) [56] |

The primary comparisons include all 1,022 QT-origin companies with known founding years through 2025, including accepted historical Possibly cases. National formation shares use worldwide foundations in the same years as the denominator. Field breadth is the number of main fields represented. Rarefaction compares the expected number of fields in equal-sized samples from the periods before and after the milestone. The sample size is the smaller of the two company counts, so a country cannot enter this comparison if either period contains no companies.

Across the twelve complete pairs, the mean increase is 7.75 companies. Its two-sided paired sign-flip p-value is 0.0156, or 0.0469 after Holm adjustment across the three endpoints. Mean field breadth increases by 1.50 fields; the equal-sample comparison increases by 0.84 fields across eight eligible pairs. Both adjusted p-values are 0.0859.

Sensitivity checks vary the time window, policy milestones and historical eligibility definition. For company counts, the Holm-adjusted sign-flip p-value is 0.0967 for two-year windows, 0.0234 for five-year windows and 0.0586 for the three-year historical-Yes-only comparison. Changing the window also changes which countries can be included. Global trends and the use of policy documents to identify companies limit what these comparisons can tell us about policy effects. Figure S3 shows the changes in national shares of worldwide company formation.

## A3 Supporting-company model

The exploratory panel covers 1999 to 2025 and the 69 jurisdictions represented by QT-origin companies with known founding years. We model supporting-company foundations using Poisson quasi-maximum likelihood with country and year fixed effects. The predictors are log2 of one plus cumulative technical-company foundations in earlier years, and the number of technical fields represented before the observation year. Technical activity covers the five non-supporting categories.

After removal of all-zero fixed-effect groups, the primary model uses 546 country-year observations across 39 jurisdictions, containing 124 supporting-company foundations. Confidence intervals use country-clustered standard errors and a t distribution with the number of country clusters minus one degrees of freedom. The two focal p-values receive Holm adjustment.

In the joint model, the rate ratio is 0.57 for a doubling of one plus prior technical foundations, conditional on field breadth, and 1.56 for one additional prior technical field, conditional on company numbers. Their unadjusted 95% intervals are 0.34–0.96 and 1.08–2.26; both Holm-adjusted p-values are 0.0392. The predictors are strongly correlated after accounting for the fixed effects ($r = 0.883$). Neither is clearly associated with supporting formation in the single-predictor specifications, and both intervals include one when the United States is excluded. The dependence on specification and strong correlation between the predictors make these associations exploratory.

## A4 Valuation sources

The historical public-company values combine disclosed share counts and daily market data. Share-count sources cover D-Wave [57, 58], Infleqtion [59, 60], IonQ [61, 62], IQM and its business combination with Real Asset Acquisition Corp. [63, 64], Quantinuum [65], Quantum Computing Inc. [66], QuantumCTek [67] and Rigetti [68, 69]. Historical listing reports provide values for QuantumCTek [70] and CIQTEK [71], while company-specific market reports cover Rigetti [72], D-Wave [73] and Quantinuum [74]. FinanceCharts provides historical market capitalisations for IonQ and Infleqtion [75]. The 16 October 2026 comparison uses Stock Analysis's reported company market capitalisations [76], counting IQM once across its listings; Yahoo Finance provides additional historical IQM data [77]. Historical currency conversions use the relevant July 2020 [78], August 2026 [79] and October 2026 [18] exchange rates. Shanghai's 16 October values are converted at the European Central Bank reference cross-rate of CNY6.6862 per US dollar [80]. The private values refer to completed financing events reported by 6 October 2026, with each company's source and valuation basis given in Table 2.

# Appendix B Additional figures

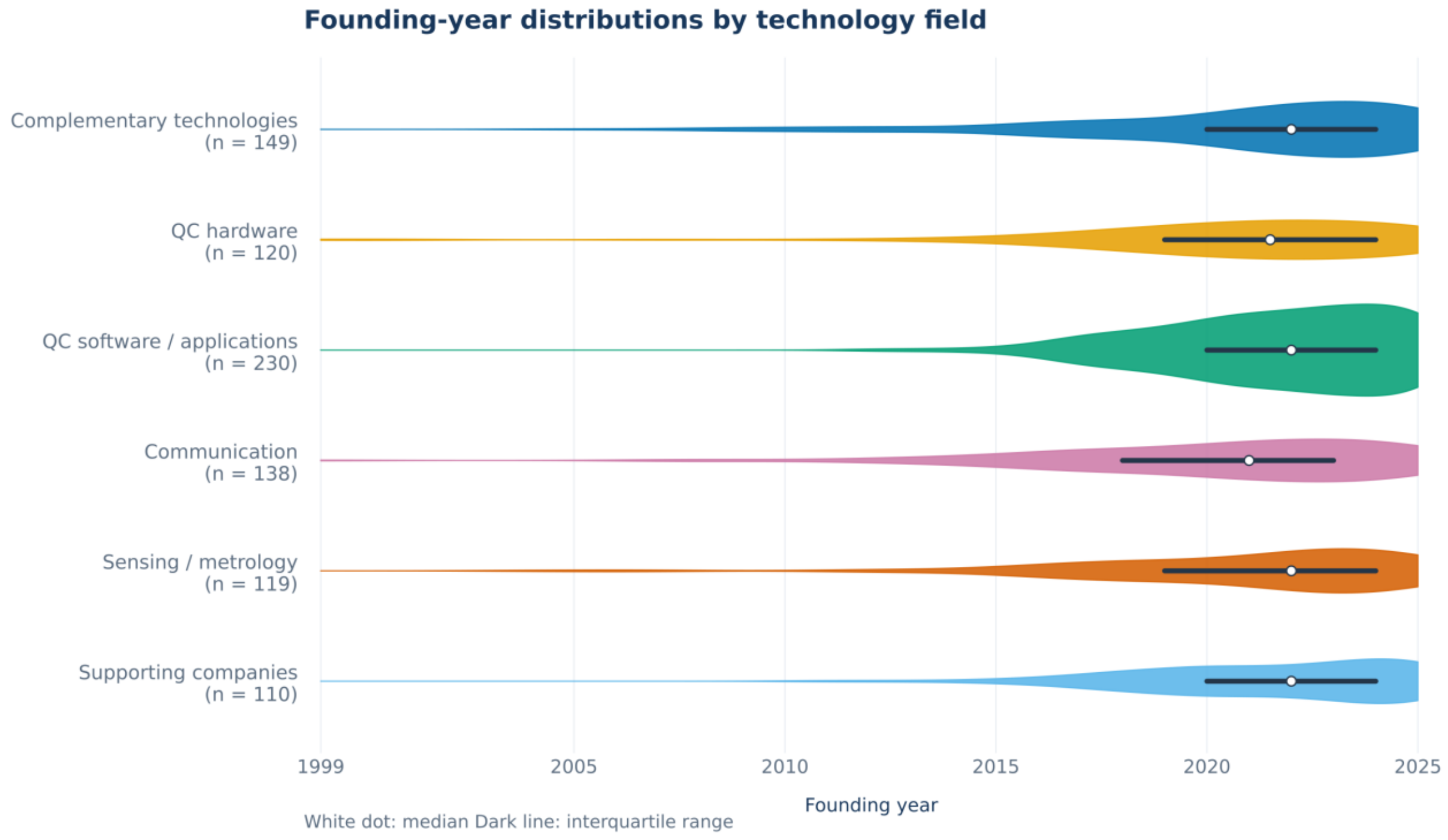

**Figure S1.** Founding-year distributions for 866 active QT-origin companies founded through 2025. The white dot shows the median and the dark line the interquartile range. Gaussian densities use Scott bandwidth and are truncated and normalised over 1999–2025; full violin areas are proportional to field totals. Width indicates temporal concentration, not an annual count.

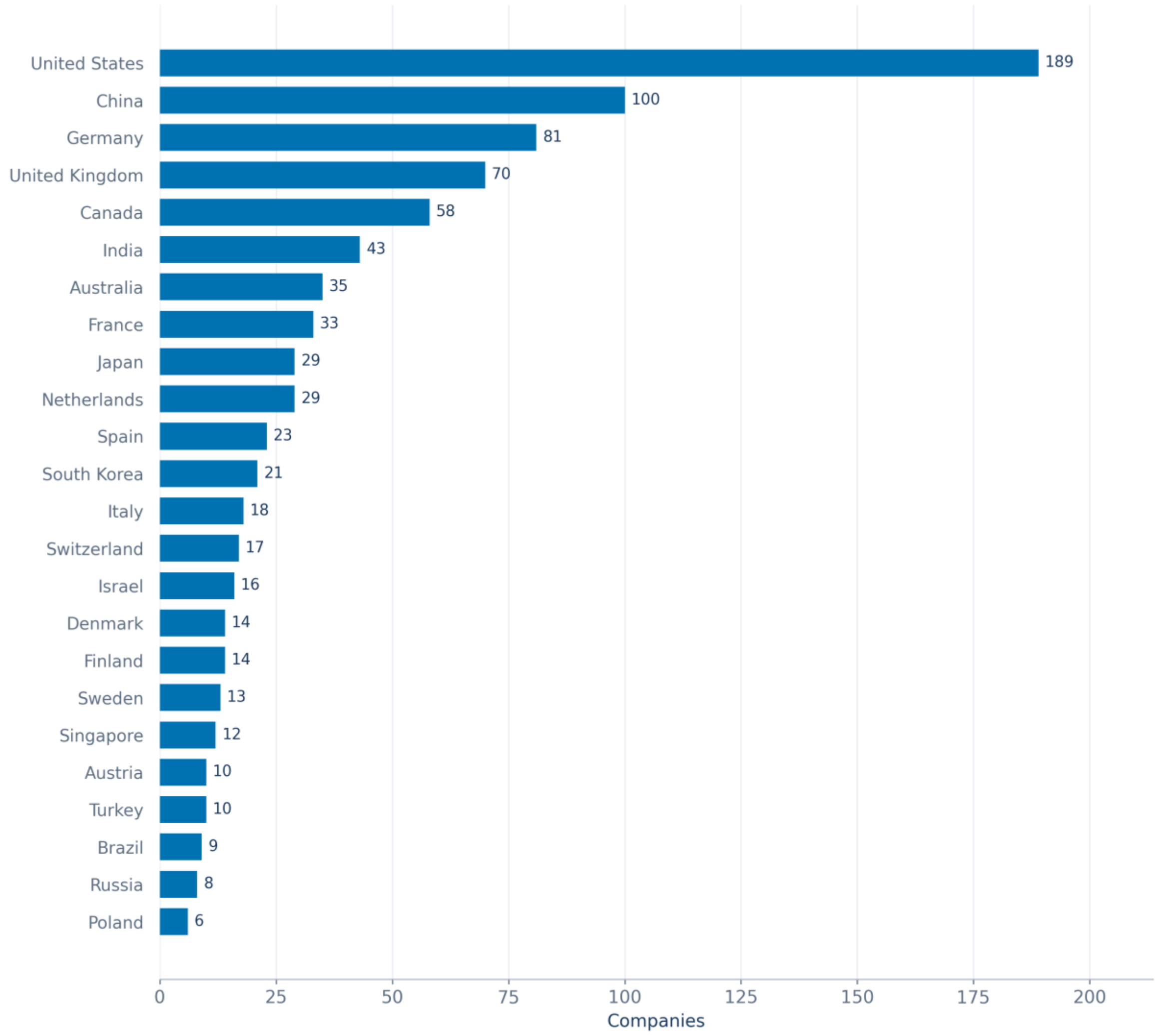

**Figure S2.** Active QT-origin companies in countries with at least five companies. Each company is counted once in its recorded country. The full population contains 935 companies across 72 countries and jurisdictions.

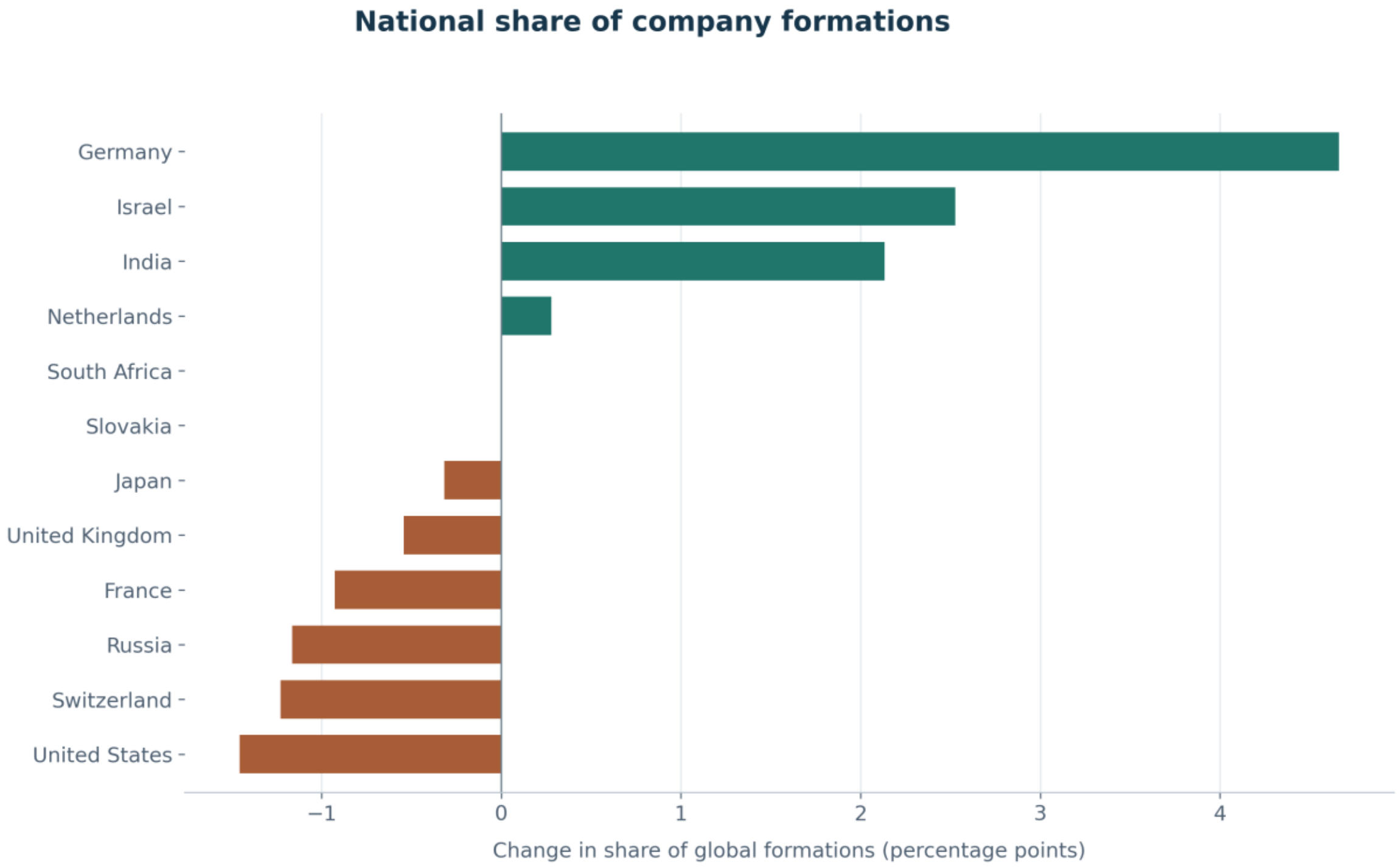

**Figure S3.** Change in each country's share of worldwide QT-origin company foundations between the three-year pre- and post-milestone windows, in percentage points. The milestone year is omitted. National and global counts use the same company cohort and years.

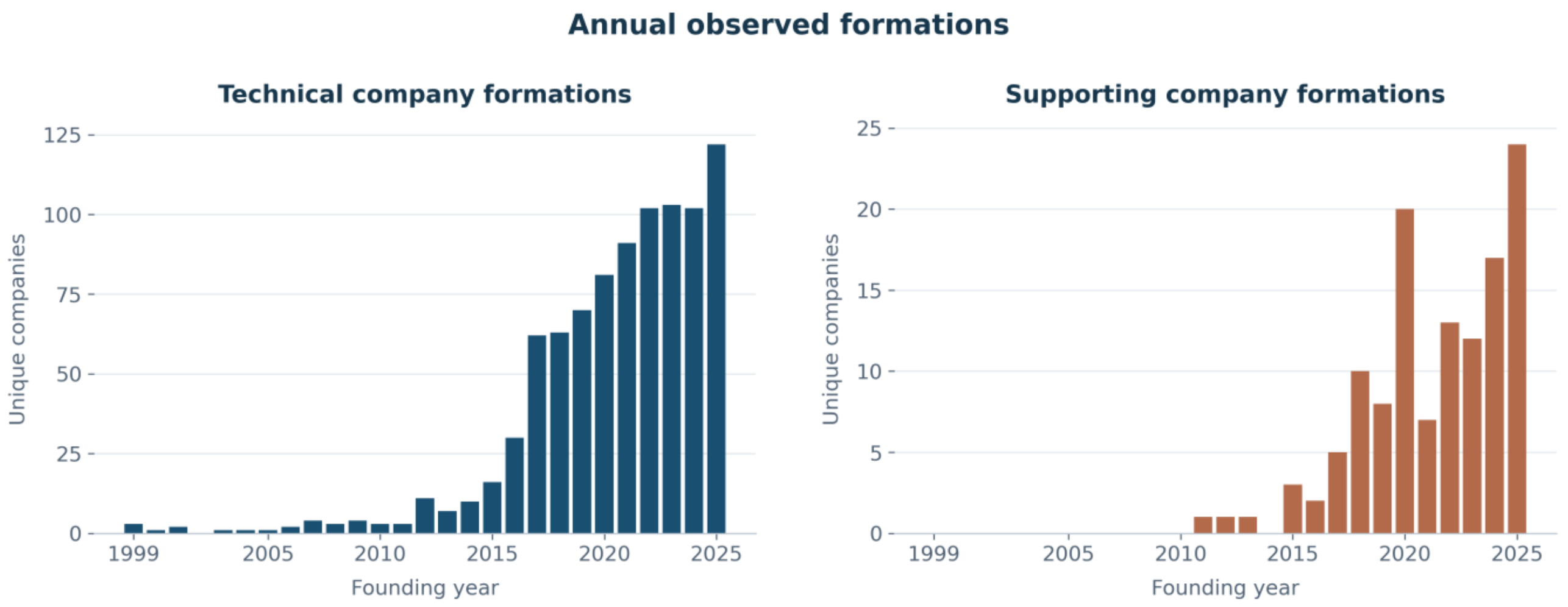

**Figure S4.** Annual foundations of 898 technical and 124 supporting companies through 2025, including active, acquired or merged, and inactive companies. Technical companies cover the five non-supporting fields, including complementary technologies. The panels use different vertical scales.

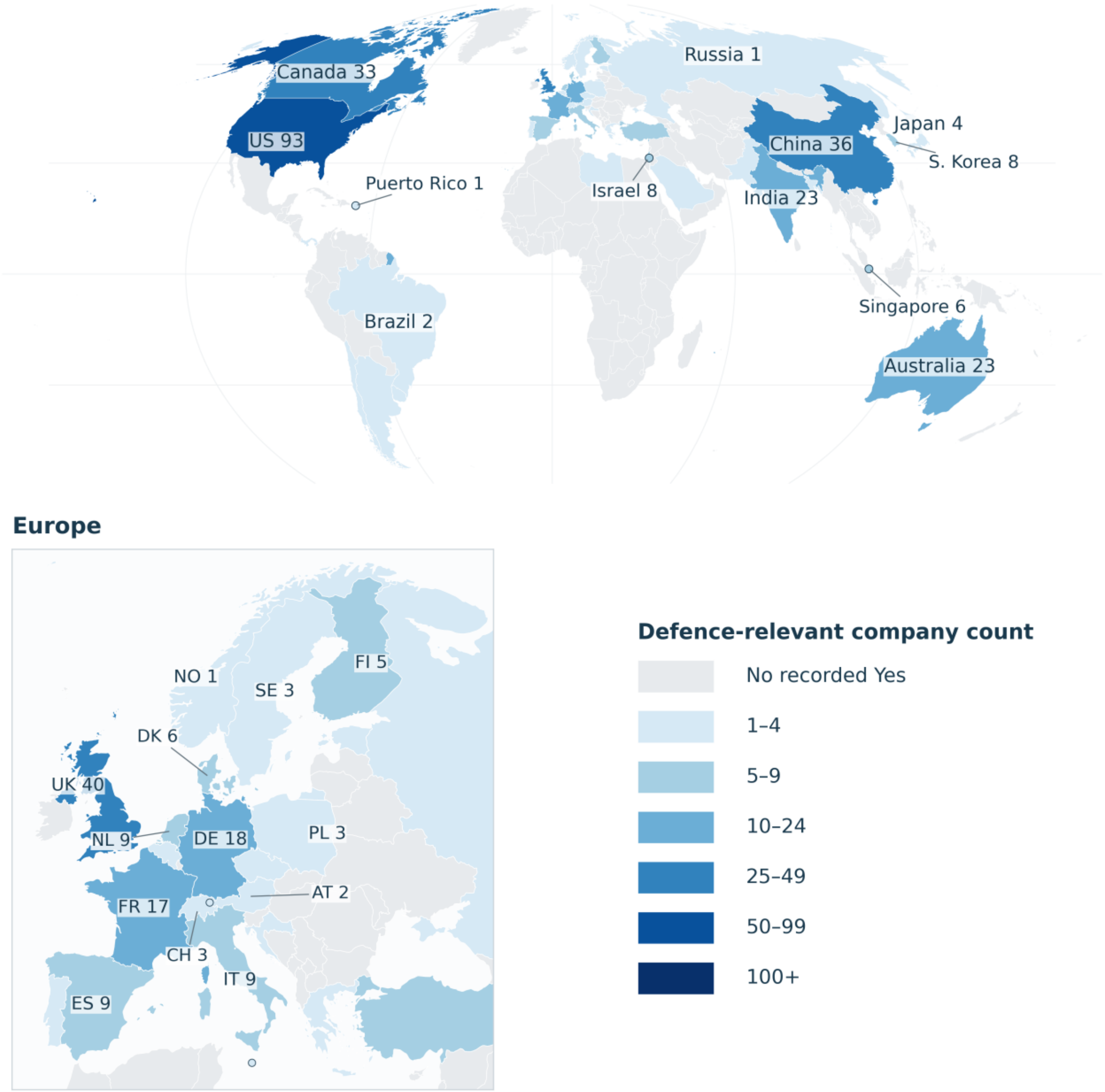

**Figure S5.** Active QT-origin companies classified Yes for current defence relevance: 390 of 935 companies, across 42 countries and jurisdictions. Grey indicates no recorded Yes-classified company. Each company is counted once in its recorded country. Boundaries: Natural Earth.

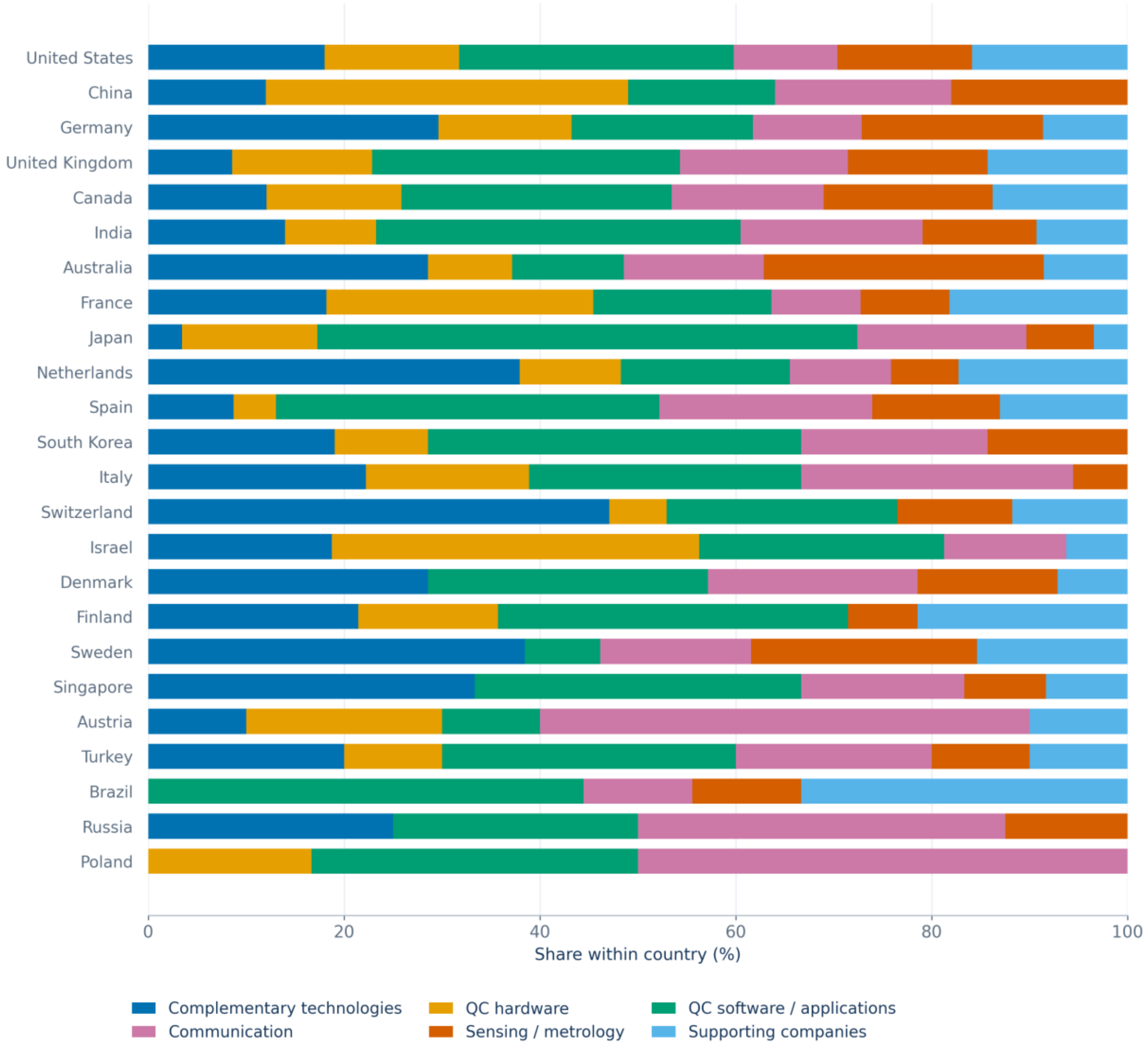

**Figure S6.** Main technology fields as percentages of active QT-origin companies within each country shown in Figure 8. Countries have at least five companies. Each company contributes one field.